\documentclass[fleqn,usenatbib]{pasa}

\usepackage{newtxtext,newtxmath}

\usepackage[T1]{fontenc}
\usepackage{ae,aecompl}

\usepackage{graphicx}	
\usepackage[space]{grffile}
\usepackage{latexsym}
\usepackage{url}
\usepackage[utf8]{inputenc}
\usepackage{fancyref}
\usepackage{textcomp}
\usepackage{longtable}
\usepackage{multirow,booktabs}
\usepackage{subfigure}
\usepackage{natbib}
\usepackage{color}
\usepackage{gensymb}
\usepackage{enumitem}
\usepackage[normalem]{ulem}

\newcommand{\msol}{M$_\odot$}

\newcommand{\OVI}{[\hbox{{\rm O}\kern 0.1em{\sc vi}}]}
\newcommand{\NV}{[\hbox{{\rm N}\kern 0.1em{\sc v}}]}
\newcommand{\CII}{[\hbox{{\rm C}\kern 0.1em{\sc ii}}]}
\newcommand{\SiIV}{[\hbox{{\rm Si}\kern 0.1em{\sc iv}}]}
\newcommand{\OIV}{[\hbox{{\rm O}\kern 0.1em{\sc iv}}]}
\newcommand{\NIV}{[\hbox{{\rm N}\kern 0.1em{\sc iv}}]}
\newcommand{\CIV}{\hbox{{\rm C}\kern 0.1em{\sc iv}}}
\newcommand{\OIII}{[\hbox{{\rm O}\kern 0.1em{\sc iii}}]}
\newcommand{\NIII}{[\hbox{{\rm N}\kern 0.1em{\sc iii}}]}
\newcommand{\AlIII}{\hbox{{\rm Al}\kern 0.1em{\sc iii}}}
\newcommand{\SiIII}{\hbox{{\rm Si}\kern 0.1em{\sc iii}}]}
\newcommand{\CIII}{\hbox{{\rm C}\kern 0.1em{\sc iii}]}}
\newcommand{\NeIV}{[\hbox{{\rm Ne}\kern 0.1em{\sc iv}}]}
\newcommand{\MgII}{\hbox{{\rm Mg}\kern 0.1em{\sc ii}}}
\newcommand{\CIIIL}{\hbox{{\rm C}\kern 0.1em{\sc iii}]\kern 0.1em{$\lambda1907,\lambda1909$}}}
\newcommand{\SII}{[\hbox{{\rm S}\kern 0.1em{\sc ii}}]}
\newcommand{\NII}{[\hbox{{\rm N}\kern 0.1em{\sc ii}}]}
\newcommand{\OII}{[\hbox{{\rm O}\kern 0.1em{\sc ii}}]}
\newcommand{\MgI}{\hbox{{\rm Mg}\kern 0.1em{\sc i}}}
\newcommand{\FeII}{\hbox{{\rm Fe}\kern 0.1em{\sc ii}}}

\newcommand{\OI}{\hbox{{\rm O}\kern 0.1em{\sc i}}}
\newcommand{\NeII}{[\hbox{{\rm Ne}\kern 0.1em{\sc ii}}] }
\newcommand{\NaI}{[\hbox{{\rm Na}\kern 0.1em{\sc i}}] }
\newcommand{\NeIII}{[\hbox{{\rm Ne}\kern 0.1em{\sc iii}}] }

\newcommand{\Halpha}{H$\alpha$}
\newcommand{\Hbeta}{H$\beta$}
\newcommand{\Hgamma}{H$\gamma$}
\newcommand{\He}{\hbox{{\rm He}\kern 0.1em{\sc ii}}}
\newcommand{\HeII}{\hbox{{\rm He}\kern 0.1em{\sc ii}\kern 0.1em{$\lambda1640$}}}
\newcommand{\HeIIoptical}{\hbox{{\rm He}\kern 0.1em{\sc ii}\kern 0.1em{$\lambda4686$} }}

\title[Massive High-Redshift Quiescent Galaxies With JWST]{Massive High-Redshift Quiescent Galaxies With JWST} 

\author[T. Nanayakkara et al.]{
Themiya Nanayakkara$^{1}$\thanks{E-mail: themiyananayakkara@gmail.com}, James Esdaile$^{1}$, Karl Glazebrook$^{1}$, Juan M. Espejo Salcedo$^{1}$, Mark Durre$^{1}$, Colin Jacobs$^{1}$
\\
\affil{$^{1}$Centre for Astrophysics and Supercomputing, Swinburne University of Technology, Hawthorn, VIC 3122, Australia.}
}

\jid{PASA}
\doi{10.1017/pas.\the\year.xxx}
\jyear{\the\year}

\usepackage{aas_macros}
\usepackage{hyperref} 
\hypersetup{colorlinks,citecolor=blue,linkcolor=blue,urlcolor=blue}

\hypersetup{draft}

\begin{document}

\begin{frontmatter}
\maketitle

\begin{abstract}
Recent ground-based deep observations of the Universe have discovered large populations of massive quiescent galaxies at $z\sim3-5$. 
With the launch of the \emph{James Webb Space Telescope (JWST)}, the on-board NIRSpec instrument will provide continuous $0.6-5.3\mu m$ spectroscopic coverage of these galaxies. 
Here we show that NIRSpec/CLEAR spectroscopy is ideal to probe the completeness of photometrically selected massive quiescent galaxies such as the ones presented by \citet{Schreiber2018}. 
Using a subset of the \citet{Schreiber2018} sample with deep Keck/MOSFIRE spectroscopy presented by \citet{Esdaile2020}, we perform a suite of mock JWST/NIRSpec observations to determine optimal observing strategies to efficiently recover the star-formation histories (SFHs), element abundances, and kinematics of these massive quiescent galaxies. 
We find that at $z\sim3$, medium resolution G235M/FL170LP NIRSpec observations could recover element abundances at an accuracy of $\sim15\%$, which is comparable to local globular clusters. 
Mimicking ZFOURGE COSMOS photometry, we perform mock spectrophotometric fitting with {\tt Prospector} to show that the overall shape of the SFHs of our mock galaxies can be recovered well, albeit with a dependency on the number of non-parametric SFH bins. 
We show that deep high-resolution G235H/FL170LP integral field spectroscopy with a $S/N\sim7$ per spaxel is required to constrain the rotational properties of our sample at $>2\sigma$ confidence. 
Thus, through optimal grism/filter choices, JWST/NIRSpec slit and integral field spectroscopy observations would provide tight constraints to galaxy evolution in the early Universe.
\end{abstract}

\end{frontmatter}



\section{Introduction}

The development of sensitive near-infrared imaging instruments such as {\tt Magellan/FourStar} \citep{Persson2013}, {\tt VLT/HAWK-I} \citep{Kissler-Patig2008},  {\tt ESO/VISTA} \citep{Sutherland2015} and {\tt UKIRT/WFCAM} \citep{Casali2007} opened a new window into the early Universe. 
Observations from these instruments detected a high abundance of red galaxies at $z\sim3-5$ \citep[e.g.][]{Marchesini2010,Spitler2014,Straatman2014,Patel2017}. 
This high abundance of massive quiescent galaxies in the early Universe posed a significant challenge to current cosmological simulations \citep{Sparre2015,Wellons2015,Dave2016,Merlin2019a}. 
The tensions between observed number densities and cosmological simulations arise due to the short evolutionary time between the Big Bang and $z\sim3-5$. 
For example, a massive quiescent galaxy with a stellar mass of $\sim2\times10^{11}$\msol\ at $z\sim3.7$ \citep[e.g][]{Glazebrook2017} needs to have formed all its stellar mass and undergone subsequent cessation of star-formation within the first $\sim1.5$ billion years of the Universe. 
Galaxy evolution and mass buildup within such a short time frame have strong implications for cosmological and chemical evolutionary models of the Universe. 
These massive $z\sim3-5$ quiescent galaxies are ideal laboratories to determine how galaxies grew and what mechanisms shut down star formation in the early Universe.

Spectroscopy of $z\sim3-5$ quiescent galaxies can be used to address the challenges faced by current cosmological models of the Universe. 
Spectroscopic confirmations of photometrically selected quiescent candidates are vital to provide tight constraints to their abundance in the $z\sim3-5$ Universe.
In addition, a detailed understanding of the stellar population properties of the $z\sim3-5$ massive quiescent galaxies and their formation mechanisms can only be obtained through the spectroscopic analysis of different elements. 
\citet{Glazebrook2017} used deep Keck/MOSFIRE observations to spectroscopically confirm the very first massive quiescent galaxy in the $z>3$ Universe. 
Their results showed that this galaxy was likely formed in a major star-formation event at $z>5$, with a star formation rate (SFR) exceeding 1000 \msol/yr, posing a significant challenge to models. 
Subsequent studies have now started to build up samples of spectroscopically confirmed massive quiescent galaxies at $z\sim3-4$ \citep[e.g.][]{Marsan2017,Schreiber2018,Tanaka2019,Forrest2020b,Forrest2020a,Valentino2020}.

Spectroscopic confirmations of several massive $z\sim3-5$ quiescent galaxies have further strengthened the need for galaxy formation models to provide efficient mass build-up and subsequent quick quenching mechanisms \citep[e.g][]{Dave2016,Merlin2019a}. 
Formation timescales of these galaxies are an important quantity to be constrained as extended formation allows the galaxy to be assembled gradually in a hierarchy of mergers relieving the tension with current models. 
Star formation history (SFH) reconstruction (from SED modelling) of three $\sim10^{11}$ \msol\ $z\sim4$ quiescent galaxies by \citet{Valentino2020} showed that the majority of the stellar mass was formed within a $\sim50$ Myr window with intense star-formation followed by an abrupt quenching event. 
SFH analysis from limited spectral features suggests average formation timescales of $\lesssim 200$ Myr, which poses a challenge to early galaxy formation models \citep{Glazebrook2017}  and hierarchical models cannot produce these massive galaxies in a single rapid event.

The formation mechanisms of these massive $z\sim3-5$ quiescent galaxies may have significant implications to the mass buildup of the early Universe. 
For example, if these galaxies were built up soon after the Big Bang in a short sharp star-formation episode followed by an abrupt quenching, they would be a fossil record of the first generation of stellar populations in the Universe. 
Studying their stellar populations would open up a unique observational window to probe the star-formation processes in the $z>6$ Universe. 
Analyzing the chemical signatures from the stars can provide vital clues to the early star-formation processes.

If these galaxies are $\alpha$-enhanced \citep[e.g][]{Kriek2016} this would suggest that the interstellar medium (ISM) was preferentially enriched by core-collapse supernovae \citep{Nomoto2006a}. 
Core collapse supernovae are end of the life products of short-lived ($\sim10-50$ Myr) massive stars  and produce $\alpha$-elements through supernovae nucleosynthesis.  
Longer lived ($>0.1$ Gyr) lower mass stars result in Type-Ia supernova and produce heavier elements such as Fe through supernova nucleosynthesis \citep{Kobayashi2009}. 
Therefore, when the star-formation episodes are shorter than $\sim1$ Gyr, the new stars formed from the gas in the enriched ISM would be $\alpha$-enhanced and would lack Fe. 
However, if the initial mass functions (IMFs) of these galaxies  vary from local values, the interpretation of the star-formation time scales becomes complicated \citep[e.g.][]{Navarro2016}.

Furthermore, these galaxies could have been built within a very short time scale from low metallicity gas with high specific star-formation rates. Under these conditions some simulations predict a preferential formation of more massive stars \citep[e.g][]{Narayanan2012}, resulting in a top-heavy initial mass function \citep[e.g][]{Chon2021a} and a higher characteristic stellar mass \citep{Sharda2021a}. 
However, currently there are no observational constraints of the stellar metallicities, $\alpha$-enhancements, or the IMF in the $z>6$ Universe. 
For example, spectral fitting to current ground-based spectroscopy of $z\sim4$ massive quiescent galaxies show degeneracies between high and low stellar metallicity solutions \citep{Saracco2020}. 
The S/N and spectral coverage of current data at $z\sim3-5$ is insufficient to develop linkages to $z\sim2$ observations of $\alpha$-enhancements \citep[e.g][]{Kriek2016}  and simultaneous $\alpha$ and Fe enhancements \cite{Jafariyazani2020}.  
Deep rest-UV continuum observations are required to constrain the stellar population properties and ISM conditions of star-forming galaxies at $z>6$, which are likely progenitors of massive quiescent galaxies at $z\sim3-5$.

The abundance, formation mechanisms, and the stellar population properties of $z\sim3-5$ massive quiescent galaxies also have strong implications for the reionization and chemical evolutionary history of the Universe. 
If the stellar populations of these galaxies were $\alpha$-enhanced and Fe deficient, the massive stars would have had less Fe blanketing and less stellar winds in the atmospheres \citep{Steidel2016}. 
Therefore, these stars would have produced higher amounts of ionizing photons compared to solar $\alpha$-abundance stars \citep{Pauldrach2001}. 
A top-heavy IMF in these galaxies could have resulted in higher rates of core-collapse supernovae.  
Core collapse supernovae have higher IMF averaged ejecta mass compared to Type Ia supernovae, which leads to stronger feedback mechanisms in galaxies \citep{Hopkins2018b}. This leads to creating more possibilities within the geometry of galaxies to create holes for ionizing photons to escape. 
The enhancement of ionizing photons along with strong supernova feedback-driven changes in the ISM geometry may alter the contribution of massive galaxies to the reionization of the Universe at $z>6$ \citep{Naidu2019}.
Thus, exploration of $z\sim3-5$ massive quiescent galaxies may lead to newer challenges in cosmology.

Kinematical properties of $z\sim3-5$ massive quiescent galaxies also provide clues for their formation scenarios. 
Deep rest-frame optical spectra have opened the door to obtaining velocity dispersion measurements of massive quiescent galaxies at $z>3$ \citep{Tanaka2019,Esdaile2020,Saracco2020}. 
When the $z>3$ quiescent galaxies are compared in the size-mass plane, they require a greater size evolution compared to what is expected from minor mergers and show evidence for dynamical masses to be lower than the stellar mass estimates \citep{Esdaile2020}. 
Additionally, despite sharing similar attributes as local elliptical galaxies, namely red colours and high stellar masses, there are indications that the morphologies of high-redshift massive quiescent galaxies are quite different. 
The modeled axis-ratio from \emph{HST} imaging shows indications of flattened ‘disc-like’ morphology in massive quiescent galaxies at $z>3$ \citep{Hill2019}.

If these galaxies are formed in short durations, mass growth through mergers would be uncommon, thus,  disc-like morphologies would be prominent at high redshift.  
However, morphological constraints at $z>3$ are poor because  existing high-resolution size measurements are based on rest-frame UV and are susceptible to underestimation from potential dust-reddening or over estimation from recent star-formation events. 
Rest-frame optical size measurements, which trace older stellar mass,  provide more robust measurements of size \citep{Kubo2018}. 
However, none of the current observations probe galaxies at sufficient depth and/or at the appropriate rest-frame wavelength windows to access features that could provide stronger constraints to the formation timescales \citep[e.g][]{Schreiber2018,Forrest2020b,Carnall2020} or morphologies.

The launch of the \emph{James Webb Space Telescope (JWST)} in 2021 provides a unique opportunity for the detailed exploration of $z\sim3-5$ massive quiescent galaxies. 
JWST Near-Infrared Spectrograph (NIRSpec) will obtain high-quality near-infra-red (NIR) spectroscopy of galaxies in the early Universe with high efficiency. 
This is due to its high sensitivity, multiplexing capabilities,  continuous spectral coverage from $0.6-5.3\  \mu \rm m$ the \citep{Birkmann2011}, low thermal background of the observatory, and the lack of atmospheric contamination in space.
Thus, by obtaining spectroscopic confirmations for photometrically selected candidates, JWST will enable astronomers to provide tighter constraints to the formation histories,  stellar populations, and kinematics of $z\sim3-5$ quiescent galaxies.

JWST/NIRSpec will obtain rest-frame optical spectra of $z\sim3-5$ quiescent galaxies,  the gold standard required to reconstruct their formation histories. 
This wavelength regime covers a variety of important absorption features which can be compared with stellar evolutionary models to determine the star-formation time scales and elemental abundances \citep{Conroy2013}. 
Balmer absorption lines have a strong dependence on the age and the SFH of galaxies \citep{Poggianti1997}. 
The variety of metal absorption lines such as Mg, Na, Ca, Ti, Na, and Fe observed in the rest-frame optical can be used with chemical evolution models to link to properties of previous star-formation episodes \citep{deLaRosa2011,Segers2016}. 
Additionally, some of these features are sensitive to different types of stars, thus they can be used to infer stellar abundances and hence the IMF of these galaxies \cite[e.g.][]{LaBarbera2013}.

In this paper, we perform a JWST/NIRSpec observational case study based on a sample of galaxies presented by \citet{Schreiber2018}.
The \citet{Schreiber2018} sample is one of the most deepest and complete spectroscopic samples of $z\sim3-5$ quiescent galaxies presented to date. 
They presented spectroscopy for 24 photometrically selected quiescent galaxy candidates with a $\sim50\%$ spectroscopic completeness rate and a $\sim90\%$ purity.  
The $z>3$ spectroscopic sample reaches $K\sim23.8$ with typical stellar masses of $\sim0.3-2.0\times10^{11}$ \msol. 
In Section \ref{sec:completeness} we discuss how JWST could be optimally used to confirm the redshift and quiescence of the photometrically selected quiescent galaxies to improve on the $\sim50\%$ completeness rate. This is a crucial first step that is necessary before investing deeper observations to obtain stellar population and dynamical properties. 
\citet{Esdaile2020} presented deep ground-based KECK/MOSFIRE spectroscopy for the brightest galaxies in the \citet{Schreiber2018} sample. 
This sample is biased towards the most massive $z\sim3-4$ massive quiescent galaxies. 
Therefore it probes an interesting range in mass, size, and velocity dispersion at $z>3$ for JWST follow-up to obtain element abundances and resolved kinematics to constrain formation scenarios of massive quiescent galaxies. 
Thus, in Section \ref{sec:stellar_pops} we discuss how JWST could enable detailed analysis of stellar population properties of the  \citet{Esdaile2020} sample and present optimal observing strategies.  
Next, in Section \ref{sec:dynamics} we discuss how JWST slit and integral field unit (IFU) spectroscopy could constrain the dynamics of the sample.  
Finally, in Section \ref{sec:discussion} we briefly discuss the expected advancements in the field with JWST and provide our conclusions. We assume a cosmology with $\rm H_0 = 70 km s^{-1} Mpc^{-1}$, $\Omega_\Lambda=0.7$, and $\Omega_M=0.3$. We use AB magnitudes throughout the paper.


\section{Probing the completeness with JWST}
\label{sec:completeness}

\subsection{The need for spectroscopic confirmations}
\label{sec:completeness_need_for_spec}

Deep multi-wavelength surveys utilizing medium-bands (such as ZFOURGE \citep{Straatman2016}) have demonstrated that galaxy Spectral Energy Distribution (SED) fitting with photometric data can obtain redshifts and evolutionary types of galaxies (i.e. blue/red star-forming or quiescent) with high-accuracy \citep[e.g.][]{Spitler2014,Straatman2014}. 
However, spectroscopy is crucial to constrain the number density of massive quiescent galaxies in the $z\sim3-5$ epoch and rule out redshift and non-quiescent outliers \citep[e.g.][]{Schreiber2018}. 
Thus, in order to make robust comparisons with cosmological models, obtaining redshift confirmations and indications for quiescence from galaxy spectra are imperative.

\begin{figure}
\includegraphics[ scale=0.8]{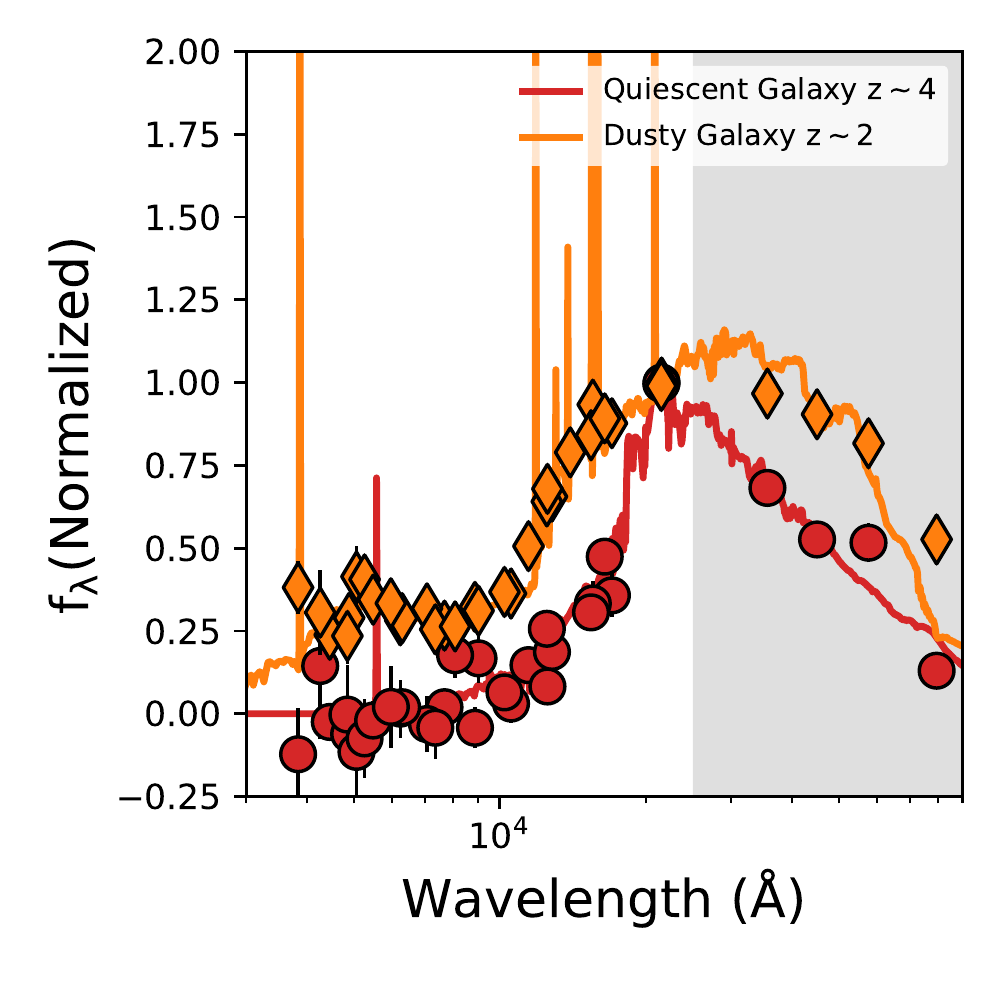}
\caption{Two example SEDs from the ZFOURGE survey. Shown in red is a quiescent galaxy SED at $z\sim4$ and in orange is a dusty star-forming galaxy at $z\sim2$. The spectra are normalized in the K band. The spectral shape of the normalized SEDs of the $z\sim4$ quiescent and $z\sim2$ dusty star-forming at $\lesssim 2\ \mu$m looks largely similar, thus spectroscopy is required to obtain emission lines to distinguish between the two types of galaxies. Alternatively, infra-red/sub-mm observations can also be used to detect the dust continuum emission of dusty star-forming galaxies. 
\label{fig:SED_QU_DU}
}
\end{figure}

The main limitation of purely characterizing quiescent galaxies based on photometry is the redshift uncertainty. 
In SED fitting techniques, the shape of the multi-wavelength photometry is used to infer the redshift. Prominent breaks such as the Lyman and Balmer breaks \citep{Pampliega2019} and the D4000 features have shown to constrain the redshift to $\sim2\%$ accuracy \citep[e.g.][]{Whitaker2011,Straatman2016}.
As an example, \citet{Straatman2016} finds that the inclusion of the {\tt FourStar} \citep{Persson2013} NIR \emph{J1, J2, J3, Hs, Hl} filters reduces the photometric redshift uncertainty by up to $\sim50\%$ at $z>1.5$.
Additionally, the redshift outlier fraction has also been shown to be dependent on the magnitude of a galaxy, with fainter sources showing a systematic increase in the outlier fraction \citep{Brinchmann2017a}.
Thus, degeneracies between redshift, galaxy type, and S/N constraints add uncertainty to the determination of photometric redshifts. 
For example, as shown by Figure \ref{fig:SED_QU_DU}, the SED shape of $z\sim2$ dusty (red) star-forming galaxies \citep[e.g][]{Spitler2014} is largely similar to that of quiescent galaxies at $z\sim4$. 
The Balmer break of $z\sim4$ galaxies observed in the $K$-band could provide some constraints, however, the uncertainties in constraining the redshift and strong \Halpha\ emission that contaminate $z\sim2$ $K$-band photometry limits the diagnostic power.   

The IR emission from the dusty star-forming galaxies can be used as an alternate indicator to differentiate them from $z\sim4$ quiescent galaxies. In the pre-JWST era, this required extremely deep Spitzer or Herschel photometry and accurate source deblending techniques to account for the large PSF in such observations \citep[e.g.][]{Stefanon2021}. 
JWST with $\sim50\times$ collecting area than Spitzer will obtain exquisite spatial quality deep near and mid-IR observations in the \emph{Hubble} legacy fields to revolutionize this space. 
Additionally, ALMA sub-mm observations can be used to detect molecular gas content \citep[e.g][]{Boogaard2021} as an indirect tracer of quiescence, though the presence of gas around the galaxy is not necessarily an indicator for ongoing star-formation \citep[e.g.][]{Kalita2021}. 
However, ALMA is not a survey instrument and thus has a limited Field of View (FoV). Therefore, covering an extragalactic field such as the \emph{Hubble} Ultra Deep Field is very expensive \citep[][]{Walter2016}.

Dusty star-forming galaxies can also be identified through rest-frame optical emission lines which can be obtained with relatively short integration times. 
Thus, spectroscopy can efficiently remove contaminants in photometrically selected galaxies such as dusty star-forming galaxies and bright active galactic nuclei.  
Spectroscopic confirmations provide tighter constraints to the evolution of the number density of massive quiescent galaxies with cosmic time at an accuracy far greater than $\sim2\%$ that can be obtained through photometric redshifts. 
Given JWST FoV this can only be achieved through spectroscopic follow up of quiescent galaxy candidates selected from  deep JWST imaging surveys \citep[e.g.][]{Kauffmann2020} and/or deep ground-based imaging surveys \citep[e.g.][]{Weaver2021}.  
Based on the ZFOURGE survey \citet{Schreiber2018} estimated the number density of quiescent galaxies between $3<z<4$ to be $\sim2\times10^{-5}$ MPc$^3$. 
The NIRCam GTO JADES survey will cover $\sim200$ arcmin$^2$ FoV reaching up to $\sim30^\mathrm{th}$ magnitude in the NIRCam filters \citep{Williams2018}. 
This is $\sim5$ magnitudes deeper than the ZFOURGE survey but only half as small in terms of the FoV. 
Therefore, JADES alone would obtain a significantly higher number of older (fainter) quiescent galaxies than any ground-based survey has currently observed.

\begin{figure}
\includegraphics[trim= 10 10 10 10 , clip, scale=0.375]{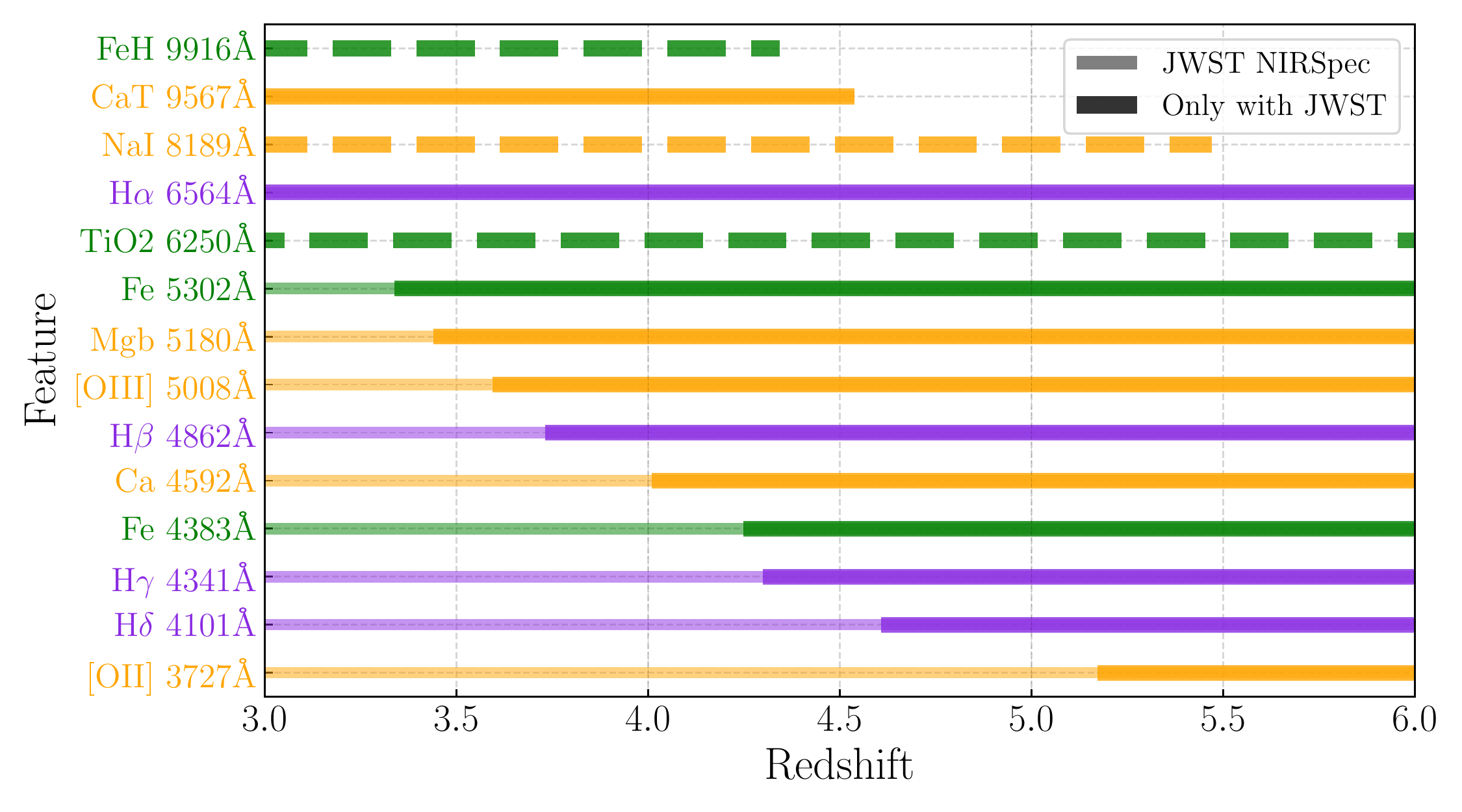}
\caption{JWST NIRSpec wavelength coverage of rest-frame optical features that are crucial to determine the quiescence and  constrain the stellar population properties of massive $z\sim3-5$ quiescent galaxies. Redshifts only accessible via space-based spectroscopy of JWST are shown by the thicker/darker color lines. Balmer emission/absorption features (sensitive to the SFH) are labeled in purple, forbidden emission lines and $\alpha$-element absorption lines are labeled in orange, while the other absorption features that constrain the overall stellar metallicity are labeled in green. IMF sensitive features are shown by dashed lines. It is evident that JWST NIRSpec spectroscopy is crucial to obtain a suite of spectral features that are necessary to analyze the $z\sim3-5$ quiescent galaxy populations. 
\label{fig:nirspec_wavelength_coverage}
}
\end{figure}

In addition to the spectroscopic confirmation of the redshift, spectroscopy provides crucial coverage of important spectral features that determines the quiescence of galaxies. 
In Figure \ref{fig:nirspec_wavelength_coverage}, we show the wavelength coverage of spectral features that are necessary to determine the quiescence and stellar population properties of $z\sim3-5$ quiescent galaxies. 
By obtaining coverage of Balmer lines or forbidden lines such as \OII\ and \OIII, strong constraints can be placed on the levels of star-formation in these massive quiescent candidates. 
Absorption features of $\alpha$-elements and other metals provide information on the stellar populations which is crucial to determine the star-formation history. 
When the redshift and quiescence is constrained through spectroscopy, multi-wavelength SED fitting would provide stronger constraints to stellar masses and SFHs, which are also important parameters to test cosmological evolution models.

\subsection{NIRSpec PRISM/CLEAR spectroscopy as a redshift confirmation machine}
\label{sec:prims_spec}

In order to address the number density of massive quiescent galaxies in the $z\sim3-5$ epoch, a systematic spectroscopic follow up of all massive quiescent galaxy candidates is required.  Rest-frame $U-V$ vs $V-J$ colour distributions have  shown to be effective in identifying the star-forming and quiescent galaxies from each other in various cosmic epochs \citep[e.g.][]{Williams2009,Straatman2014}. 
Recent ground-based efforts have been successful in obtaining the rest-frame optical coverage of the brightest of these quiescent candidates \citep[e.g.][]{Schreiber2018,Valentino2020}. 
They have shown a high success rate in identifying quiescent galaxies based on rest-frame  $U-V$ vs $V-J$ colours with a high purity but not necessarily high completeness. For example \citet{Schreiber2018} analyzed 24  $U-V$ vs $V-J$ colour selected galaxies and obtained redshift confirmations for 12 galaxies out of which only two were $z\sim2$ dusty star-forming interlopers (see discussion in \citet{Schreiber2018}).

However, studies have so far only targeted the brightest quiescent galaxies in the $z\sim3-5$ epoch. 
This is driven by observational challenges in obtaining ground-based NIR spectroscopy. The limited wavelength coverage due to atmospheric absorption,  strong skyline contamination, along with limited multiplexing capabilities in ground-based NIR instruments have traditionally challenged astronomers to spectroscopically follow up mass/magnitude complete samples of quiescent galaxies. 
Additionally, targeting only the brightest galaxies biases the observed samples since most massive quiescent galaxies are in general older and fainter.

The multiplexing capability and the $0.6-5.3\mu \rm  m$ wavelength coverage makes JWST/NIRSpec an ideal instrument to obtain spectroscopic redshifts of photometrically selected massive quiescent galaxy candidates. 
The PRISM/CLEAR disperser/filter combination of NIRSpec provides continuous spectral coverage between $0.6-5.3\ \mu \rm  m$ at a resolution of $R\sim100$. 
Thus, by selecting quiescent candidates from deep ground and space-based imaging surveys, NIRSpec PRISM/CLEAR spectroscopy can be used to obtain the crucial rest-frame optical absorption features that are necessary to confirm the redshifts, quiescence, and rule out interlopers. 

\begin{figure}
\includegraphics[trim= 10 10 10 10 , clip,  scale=0.35]{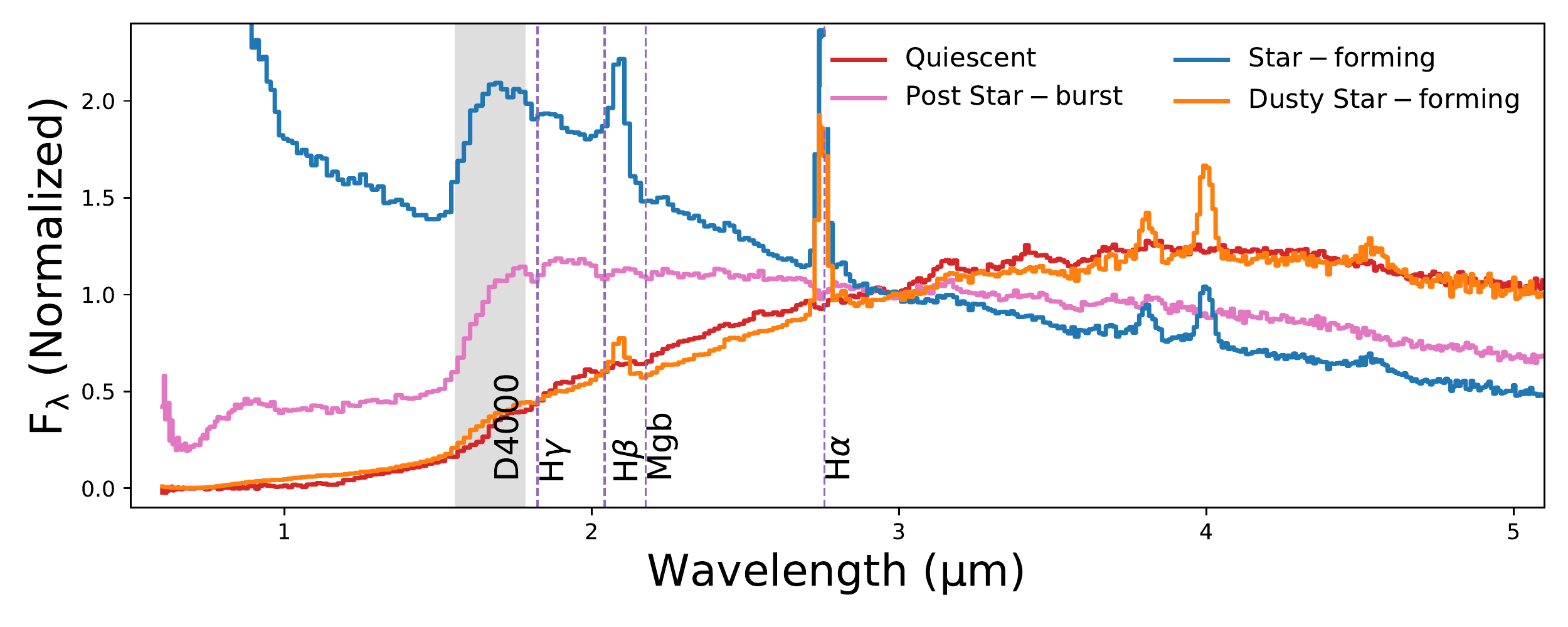}
\caption{Simulated observations of JWST/NIRSpec PRISM/CLEAR spectroscopy for a quiescent $(E_{(B-V)}=0.5)$, a post star-burst $(E_{(B-V)}=0.3)$, dusty star-forming $(E_{(B-V)}=1.0)$, and dust-free star-forming ($(E_{(B-V)}=0.0)$)  galaxy at $z\sim3$. Redder NIR colours due to age-dust degeneracies make it challenging to identify quiescent galaxies purely based on photometric observations. The high sensitivity coupled with the multiplexing capability makes NIRSpec an ideal instrument to obtain $0.6\mu m-5.3\mu m$ spectroscopy to accurately distinguish between these types of galaxies and confirm their quiescence. The spectra shown here are all normalized to $K=21.5$ and the typical continuum S/N obtained is $\sim80-100$ in a $\sim1500$ s exposure.
\label{fig:prism_spectra}
}
\end{figure}

In Figure \ref{fig:prism_spectra} we show that the NIRSpec PRISM/CLEAR spectroscopy is capable of accurately distinguishing between $z\sim3$ quiescent, post-starburst, dusty (red) star-forming, and dust-free (blue) star-forming galaxies. 
Blue star-forming galaxies can be clearly identified based on the NIR colours and rest-UV continuum, however, prism spectra provide emission line measurements to further constrain the ISM and star-formation properties of these galaxies.  
NIR colours will be similar between dusty star-forming and quiescent galaxies, however, the Balmer and forbidden emission features in the star-forming galaxies will be detectable by $R\sim100$ spectroscopy to rule out any star-forming contaminants. 
\Hbeta\ and \Hgamma\ equivalent width can be used to determine the time since the last star-formation episode \citep[e.g.][]{Glazebrook2017} and comparisons of the continuum around D4000 feature and the rest-frame near-UV continuum can distinguish between post-starburst and dusty star-forming/quiescent galaxies.

In addition to the simultaneous $\rm 0.6-5.3\mu m$ wavelength coverage, the sensitivity of the NIRSpec instrument and the lower background level in space-based observations increases the efficiency of the JWST observations.  
Spectra shown in Figure \ref{fig:prism_spectra} are generated using the Flexible Stellar Population Synthesis (FSPS) code \citep{Conroy2010b} and are normalized to $K=21.5$ in the JWST exposure time calculator (ETC). 
With an exposure time of $\sim1500$s, ETC predicts that NIRSpec PRISM/CLEAR spectroscopy will obtain a typical continuum S/N of $\sim80-100$ for these galaxies. 
Dusty star-forming, post-starburst, and quiescent galaxies are reddened with a $E_{(B-V)}=1.0, 0.3, 0.5$ respectively, following the \citep{Cardelli1989} extinction law. 
No extinction is applied to the dust-free starforming galaxies.
For simplicity, we assume the sources to be pointlike. 
\citet{Glazebrook2017} obtained a S/N of 6 for a $K=22.5$ $z=3.7$ quiescent galaxy with 7 hours of $K-$band Keck/MOSFIRE spectroscopy at a resolution of 19 \AA\ per pixel. For a $K=21.5$ source Keck/MOSFIRE can reach a S/N of $\sim9$ in 1500 s at similar resolution to NIRSpec PRISM/CLEAR spectroscopy\footnote{Such observations will be carried out by JWST Cycle 1 General Observers program ID 2565	\emph{``How Many Quiescent Galaxies Are There at $3<z<4$ Really?''}.}. 
Therefore, even when only a limited wavelength coverage is considered, NIRSpec PRISM/CLEAR spectroscopy is $\sim\times10$ more efficient.

As a potential caveat to the efficiency of quiescent galaxy surveys using NIRSpec PRISM/CLEAR, we also note that the NIRSpec micro-shutter assembly (MSA) adds extra observational challenges in obtaining uncontaminated spectra of galaxies. In addition to contamination from fixed open shutters, as stated in the JDox, MSA flux leakage due to small gaps in the MSA shutters can lead to $\sim2\%$ contamination\footnote{https://jwst-docs.stsci.edu/near-infrared-spectrograph/nirspec-instrumentation/nirspec-micro-shutter-assembly\#NIRSpecMicro-ShutterAssembly-msa\_leakageMSAfluxleakage}.  
Therefore, to obtain deep spectroscopy of faint spectral features, the NIRSpec Fixed Slits (S200A1 or S200A2 with 0.2'' slit width) is ideally suited.  Fixed Slits provide the  cleanest spectra in the NIRSpec detectors and yield a higher S/N compared to the MSA mode for the same exposure time. 
Additionally, it is possible to obtain simultaneous MSA and fixed slit spectroscopy by positioning the primary target of interest in the fixed slit location and populating the MSA shutters with secondary targets. This can be achieved by either configuring the MSA to only allow for rotational variation using the MSA Planning design tool or by manually configuring the MSA shutters in open/close positions.

\section{Stellar Populations with JWST}
\label{sec:stellar_pops}


\subsection{Stellar population models used in this analysis}
\label{sec:stellar_pop_models}

We use two different stellar population models in our analysis. 
To investigate how the spectral features change based on element abundances we use the  \citet{Villaume2017} empirical SSPs using the {\tt alf} software \citep{Conroy2018a} to generate synthetic galaxy spectra.  
\citet{Villaume2017} models have the flexibility to change the abundance of individual elements which makes it ideally suited for this task. 
We use the same models to also investigate the accuracy in recovering element abundances from mock observed spectra as detailed in Section  \ref{sec:abundances}.

\begin{figure}
\includegraphics[ scale=0.8]{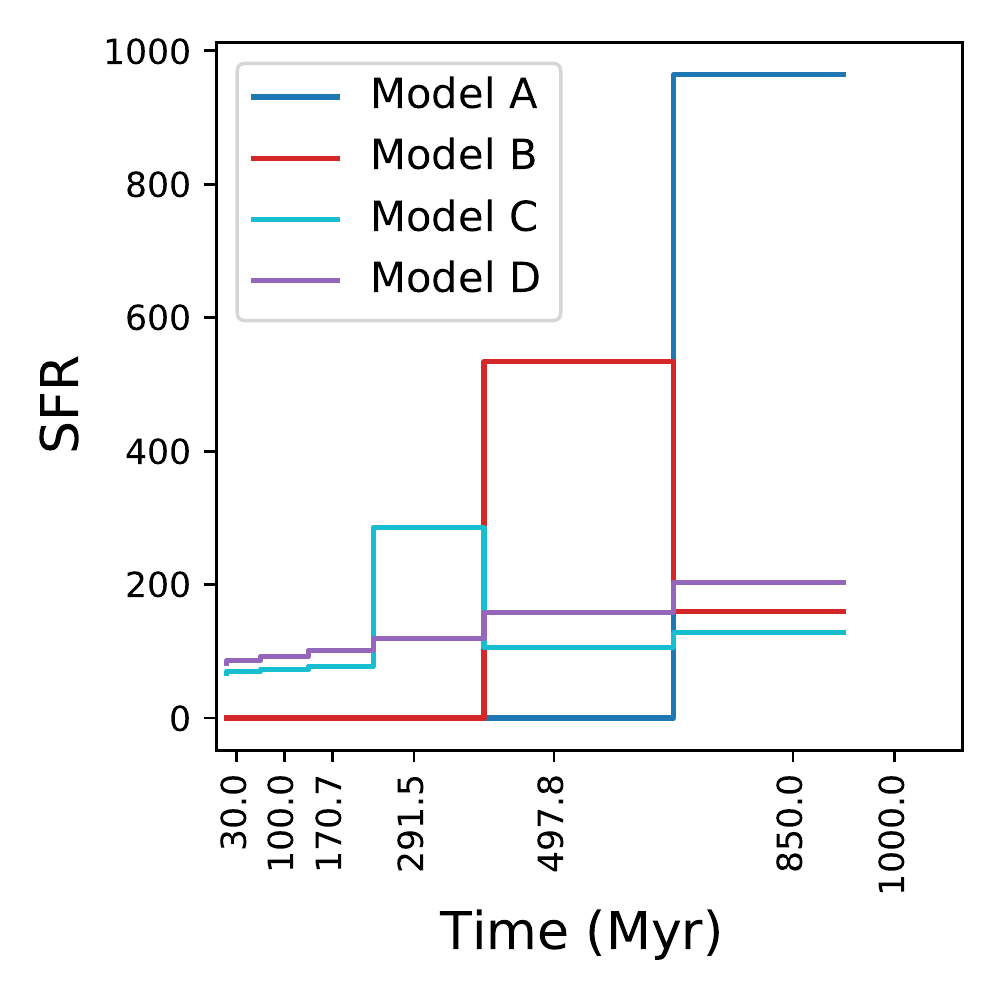}
\caption{Visualization of input SFHs generated using FSPS models as detailed in Table \ref{tab:sfh_properties}. The input parametric forms are averaged using 7 time-bins which are fixed similar to the {\tt Prospector} fits as described in Section \ref{sec:sfh}. In lookback time, the first two bins are fixed to be between 0-30 Myr and 30-100 Myr. The most distant bin is fixed to be between 850-1000 Myr. The remaining 4 bins are split in logarithmic time evenly between 100 Myr and 850 Myr.
\label{fig:input_sfhs}
}
\end{figure}

To investigate the role of SFH in recovering galaxy properties we use the FSPS code \citep{Conroy2010b} to generate synthetic spectra for 4 mock  galaxies as described by Table \ref{tab:sfh_properties}.
Within FSPS we select the MILES spectral library \citep{Vazdekis2010} and MIST isochrones \citep{Choi2016} to generate the synthetic models. 
All galaxies are generated at solar metallicity following a \citet{Calzetti2000} dust law with an optical depth of $\tau_{dust}=0.5$. Unless specified all other parameters are kept at the {\tt python-fsps} default values.
Galaxy Model A is formed via a single burst event at $t=0$.
Model B has a delayed $\tau$ model in the form of $te^{-t/\tau}$ with $\tau=1 Gyr$. A constant SFH is superimposed in the model to generate 50\% of the total mass at 1 Gyr and the star-formation is truncated at 500 Myr.
Model C is an exponentially declining SFH with a  timescale of $\tau=1$ Gyr. 20\% of the total mass at 1 Gyr is formed through a constant SFH episode. Both of these are combined with an instantaneous burst at 700 Myr that generates 30\% of the total stellar mass at 1 Gyr.  
Model D is an exponentially declining SFH with a timescale of $\tau=1$ Gyr. 
Model C and D have declining but residual star-formation while Models A and B have been quenched for $>0.5$ Gyr.  
All FSPS SFH parameters used are presented in Table  \ref{tab:sfh_properties} and are also shown by Figure \ref{fig:input_sfhs}.

The FSPS galaxies are normalized to a $10^{11}$ \msol\ to  be similar to the \citet{Esdaile2020} sample. 
Observed galaxy spectra are generated at 1 Gyr and are fed through the JWST ETC to compute the observed spectrum from the G235M/FL170LP grism/filter combination with a continuum S/N$\sim30$ per pixel. ETC output spectra are instrument throughput corrected and flux calibrated.

\begin{table*}
\caption{SFHs generated using FSPS models. The parameters related to the SFH used in FSPS are shown by the columns. Model A is an SSP while models B, C, and D have parametric SFHs. All models are generated at solar metallicity at an age of 1 Gyr.}
\centering
\begin{tabular}{@{}cccccccc@{}}
\hline\hline
 Model Name  & {\tt sfh} & {\tt tau} & {\tt const} & {\tt sf\_start} & {\tt sf\_trunc} & {\tt tburst} & {\tt fburst}  \\ 
\hline%
A 			& 0 		& 	--		& 	--  		& 		--		& 		--			& 		--		&  	--			\\
B 			& 1 		& 	1.0		& 	0.5			& 		0.0		& 		0.5			& 		0.0		&  	0.0			\\
C 			& 1 		& 	1.0		& 	0.2			& 		0.0		& 		0.0			& 		0.7		&  	0.3			\\
D 			& 1 		& 	1.0		& 	0.0			& 		0.0		& 		0.0			& 		0.0		&  	0.0			\\
\hline\hline
\end{tabular}
\label{tab:sfh_properties}
\end{table*}

\subsection{Element abundances}
\label{sec:abundances}

\begin{figure*}
\includegraphics[ scale=0.8]{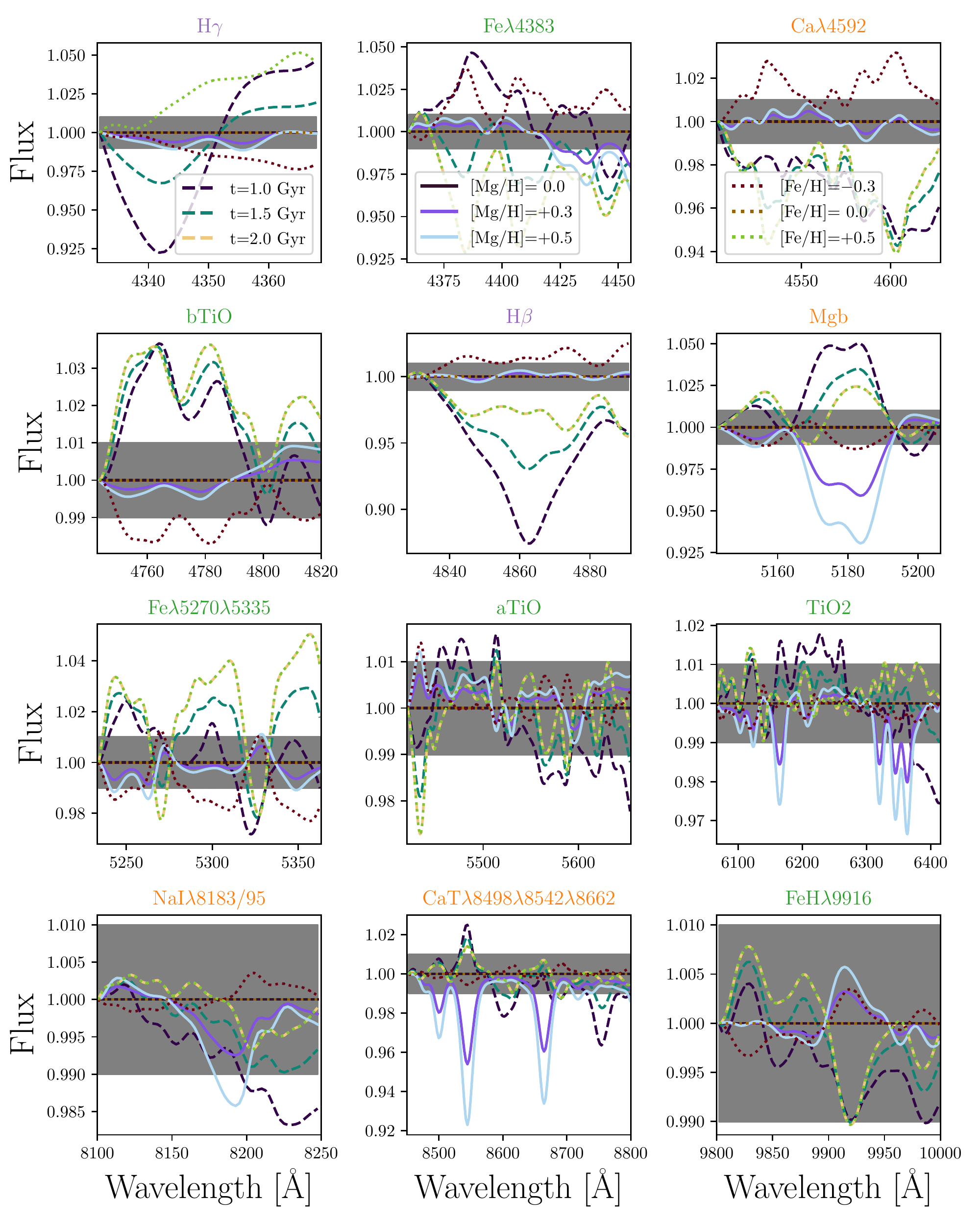}
\caption{Here we show \citet{Villaume2017} empirical SSPs computed at different ages (dashed), $\alpha$-abundances (solid), and metallicities (dotted) using {\tt alf}. Spectra are smoothed to a resolution of 100 km/s and are divided by a 1.5 Gyr old solar abundance spectrum, so relative changes in the spectra can be clearly identified.  The grey shaded region shows the relative accuracy that is obtained by a S/N=100 spectrum. It is evident that age through the Balmer absorption lines, $\alpha$-abundances through Mgb, and metallicity ([Fe/H]) through Fe features can be recovered using these individual absorption features at this S/N level.  
\label{fig:alf_features_abun}
}
\end{figure*}

Stellar population properties of galaxies can be obtained via rest-UV and optical absorption line spectroscopy. 
O and B-type stars which contribute to UV flux have short lifetimes. 
Therefore,  rest-frame optical features from A and G type stars play the most important role in deciphering the underlying stellar population properties of quiescent galaxies. 
In Figure \ref{fig:alf_features_abun}, we show prominent rest-frame optical absorption lines that are necessary to constrain the ages, metallicities, and element abundances of quiescent galaxies based on the \citet{Villaume2017} empirical SSPs. 
We choose SSPs purely for illustrative purposes because it simplifies the variation of the selected spectral features with time, $\alpha$-abundance, and metallicity. 
Realistic SFHs with multiple generations of stellar populations require more advanced simultaneous treatment of multiple spectral features  and are further explored later in the context of full spectral fitting.

It is clear from  Figure \ref{fig:alf_features_abun} that not all features differ substantially between models such that they can be distinguished at a given sensitivity.
As an example, the SFH (or rather the time since the last star formation episode) can be well constrained by the Balmer absorption lines with reasonable S/N limits ($\sim100$). 
However, in order to map changes in the lower-mass IMF slope through the IMF sensitive features such as the Fe-H band and NaD absorption \citep[e.g.][]{ConroyVanDokkum2012} requires extremely high S/N sensitivity of $\gtrsim350$. Achieving such levels of S/N in spectroscopy for populations of galaxies at $z>3$, even with JWST requires upwards of hundreds of hours per target and is not practical. 
However, we note that these S/N levels are only presented in the context of simple SFHs and assume that diagnostics rely on a single feature. Much tighter constraints can be made by combined analysis of spectral features. Local galaxy analysis between individual features (i.e. Lick indices \citep{Burstein1984}) and full spectral fitting have shown that results from both methods are in broad agreement with each other \citep[e.g][]{Conroy2014}.

Next, we go beyond individual spectral line analysis and perform full spectrum fits to identify optimal S/N levels and JWST/NIRSpec grism/filter combinations required to constrain the element abundances and the SFHs of $z\gtrsim3$ quiescent galaxies similar to that of \citet{Esdaile2020} sample.    
The combined role of spectroscopy and multi-wavelength photometry in constraining the complex SFH of galaxies will be addressed in Section \ref{sec:sfh}.

Rest-frame optical spectroscopy covers the spectral features that are necessary to constrain the formation history of quiescent galaxies. In addition to constraining the last star-formation episode from the Balmer absorption features, the coverage of absorption features from elements such as Mg, Fe, Ti provides additional constraints to the durations of previous star-formation episodes, metallicities, and chemical abundance patterns of galaxies \citep[e.g][]{Vazdekis2010}. 
These are crucial to determine the SFHs of the early quiescent galaxies and can be compared with $\Lambda$CDM hierarchical galaxy formation models.

Here, we investigate what optimal JWST NIRSpec grism/filter combinations and S/N thresholds are required to accurately obtain element abundances of a simulated massive quiescent galaxy similar to that of \citet{Esdaile2020} sample. 
We use \citet{Villaume2017} empirical SSPs to generate a mock 1.5 Gyr old galaxy at $z=3.2$ with velocity dispersion of $300$ km/s and keep all elemental abundances at solar. 
We feed the mock galaxy to the JWST ETC to obtain a suite of mock observables at different S/N values for the JWST NIRSpec fixed slit S200A1 with G235H/FL170LP and G395H/FL290LP grism/filter combinations. 
We then use the full spectral fitting code {\tt alf} \citep{Conroy2018a} on the calibrated data to fit for velocity, velocity dispersion, age, [Z/H], and element abundances of C, N, O, Mg, Si, Ca, Ti, and Na. 
Spectral fitting is performed to each individual grism/filter combination and to the combined spectrum from both grisms/filters to obtain 100 best-fit values for the input spectrum.

\begin{figure*}
\includegraphics[ scale=0.8]{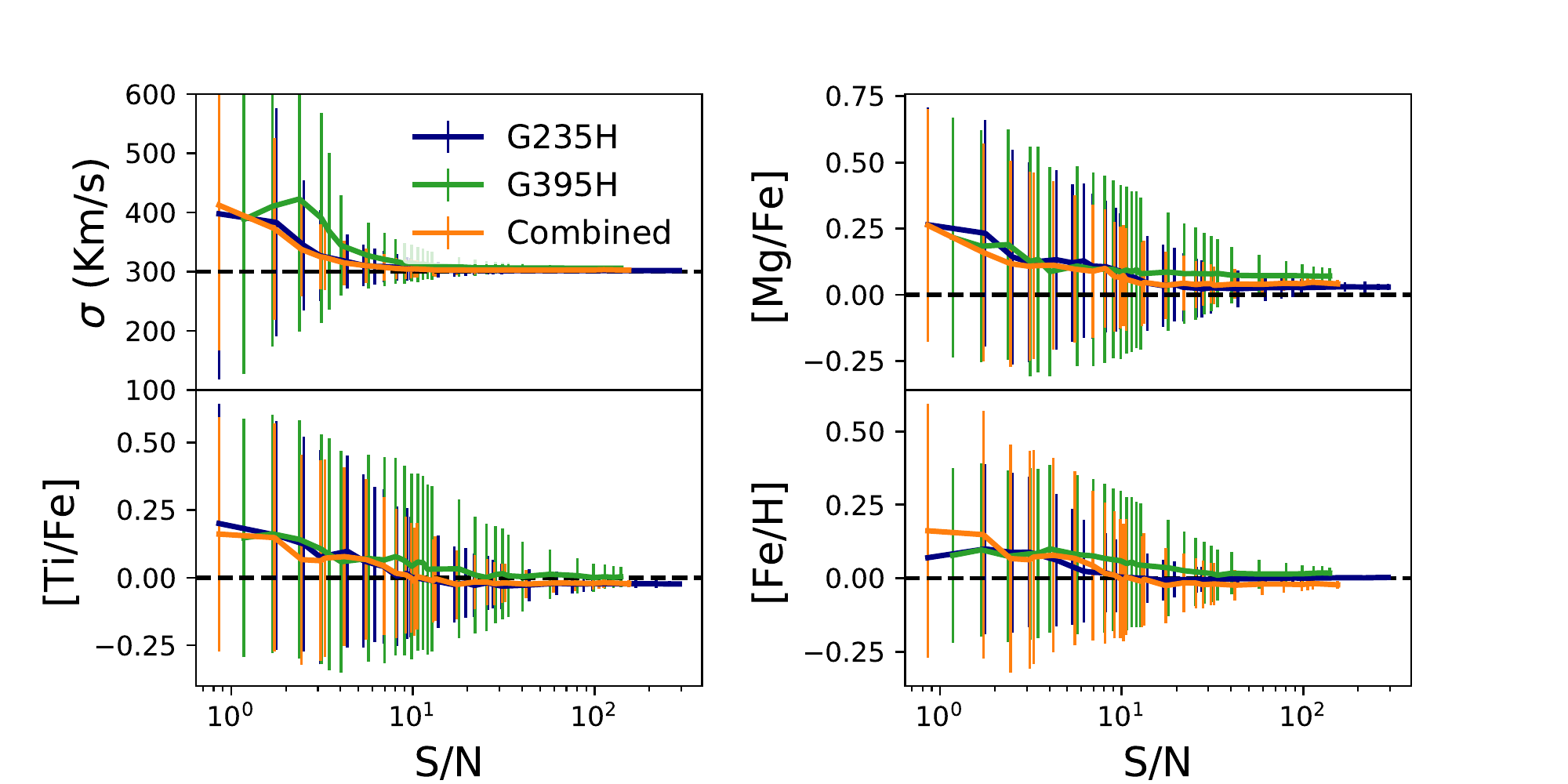}
\caption{The recovery of {\bf Top Left:} velocity dispersion and elemental abundances of {\bf Top Right:} [Mg/Fe], {\bf Lower Left:} [Ti/Fe], {\bf Lower Right:} [Fe/H] of mock JWST NIRSpec S200A1 G235H/FL170LP and G395H/FL290LP observations. Full spectral fitting is performed using {\tt alf} for individual grism/filter combinations separately and together at their respective native grism resolutions. The input values to the model spectra are shown by the horizontal dashed lines. 
\label{fig:alf_recoveries_GXXXH}
}
\end{figure*}

In Figure \ref{fig:alf_recoveries_GXXXH} we show the element abundance recovery of [Mg/Fe], [Ti/Fe], and [Fe/H] for our mock observables. It is evident that at native resolutions of each grism/filter combinations, at a S/N$\gtrsim30$ per pixel the recovery of the elemental abundances converges to the input values. We further find that there is no significant difference between the two grism/filter combinations. However, when the spectra from both grism/filter combinations are fit together, the uncertainty is slightly decreased. 
Even though spectral features such as Mgb are not covered by G395H/FL290LP for $z=3.2$ galaxies, inherent relationships between various $\alpha$-elements and other metals in the \citet{Villaume2017} templates allows Mg abundance to be converged albeit with slight systemic offset. 
For example, because Mg and Ca are both $\alpha$ elements, the coverage of the CaT$\lambda8498\lambda8542\lambda8662$ features can be used to constrain the Mg abundance using theoretical response functions \citep{Conroy2018a}.  
Thus, abundances of elements not included in the spectral range are inferred from predetermined element abundance ratios and chemical evolutionary models. However, direct measurements of element abundances should be preferred particularly for high redshift galaxies where these assumptions may not hold.

\begin{figure*}
\includegraphics[ scale=0.8]{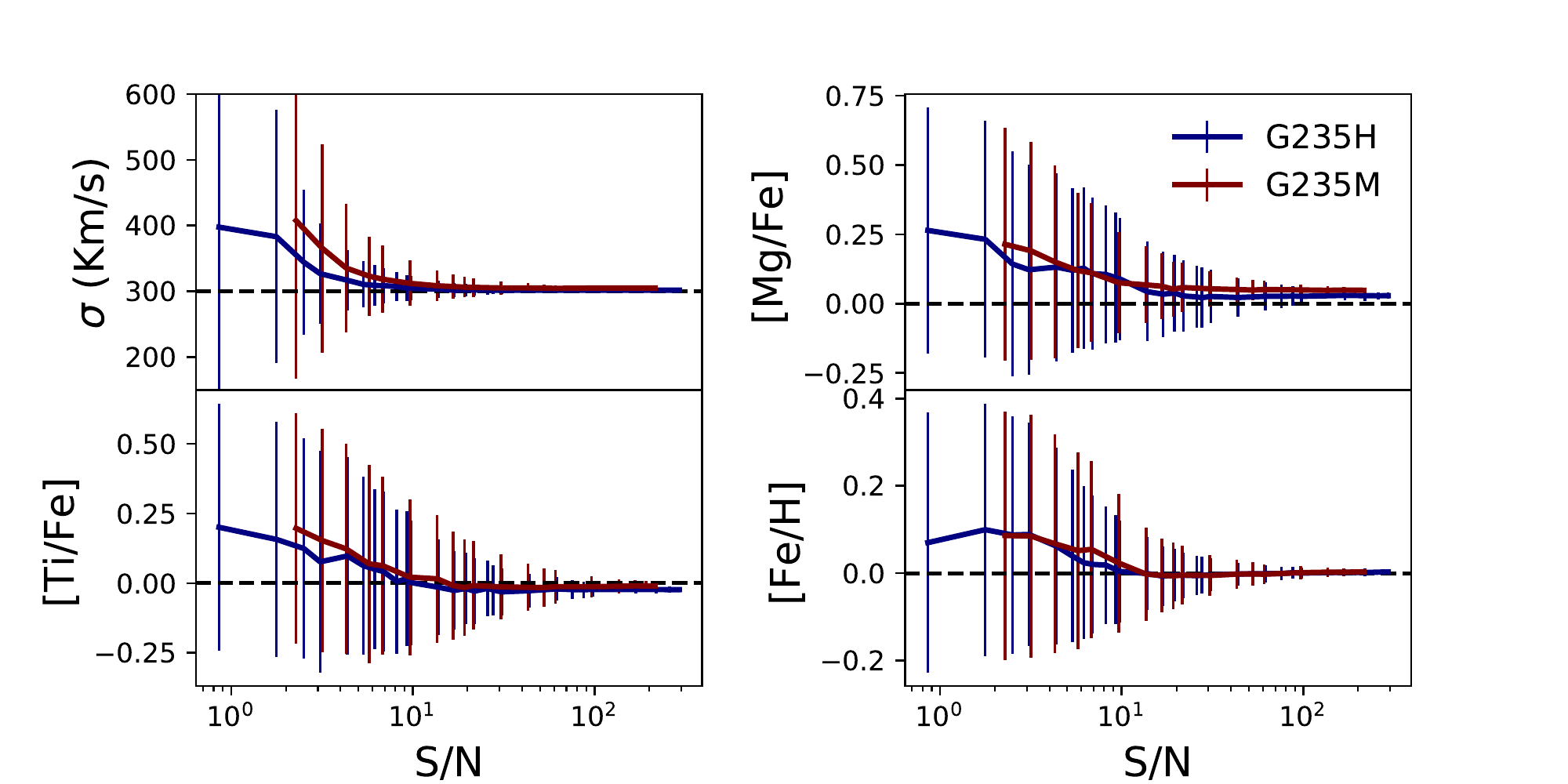}
\caption{The recovery of {\bf Top Left:} velocity dispersion and elemental abundances of {\bf Top Right:} [Mg/Fe], {\bf Lower Left:} [Ti/Fe], {\bf Lower Right:} [Fe/H]  from NIRSpec G235M/FL170LP and G235H/FL170LP grism/filter combinations using {\tt alf} at their respective native grism resolutions. The true value is shown by the horizontal dashed lines. It is evident that even with the lower resolution G235M grism, the input parameters can be accurately recovered at S/N$\gtrsim30$. For the same exposure time the S/N can increase by a factor of $\gtrsim2$ between G235H and G235M grisms, therefore, obtaining absorption line spectroscopy using the G235M will be the most efficient to derive velocity dispersions and element abundances. 
\label{fig:alf_recoveries_G235}
}
\end{figure*}

Thus, for the $z\sim3-4$ \citet{Esdaile2020} sample, most of the spectral features that are necessary to obtain $\alpha$-element abundances can be obtained by the G235H/FL170LP grism/filter combination. 
For the fixed observed wavelength coverage of the FL170LP filter, we investigate whether the choice of the spectral resolution offered by the G235M and G235H grisms play a significant role in the convergence of mock observable parameters. 
In Figure \ref{fig:alf_recoveries_G235} we show the recovery of the parameters for the G235M and G235H grisms.
It is evident that there is no significant difference between the two grisms for the recovery of the [Mg/Fe], [Ti/Fe], and [Fe/H] element abundances. Therefore, it is advantageous to obtain G235M spectroscopy to increase the efficiency of $z>3$ quiescent galaxy observing programs.  
From our {\tt alf} simulations we find that at a S/N of $\sim30$ per pixel, an accuracy of $\sim15\%$ can be obtained for element abundances which are comparable to the accuracy obtained for local globular clusters \citep{Conroy2018a}.

Next, we go beyond SSPs to investigate the role of SFH in recovering element abundances from  the G235H/FL170LP grism/filter combination. 
We use FSPS code \citep{Conroy2010b} to generate 4 different SFHs as detailed in Section \ref{sec:stellar_pops}.  
Simulated galaxy spectra are generated at 1 Gyr of age with a velocity dispersion ($\sigma$) of 300 km/s.
We then use the JWST ETC to generate mock observed spectra at $z=3.2$ using the  NIRSpec G235M/FL170LP grism/filter combination. The galaxy spectra are normalized to a stellar mass of $10^{11}$\ \msol\ at 1 Gyr and the exposure parameters are varied such that a continuum S/N of $\sim30$ is achieved for the spectra.

\begin{figure*}
\includegraphics[ scale=0.8]{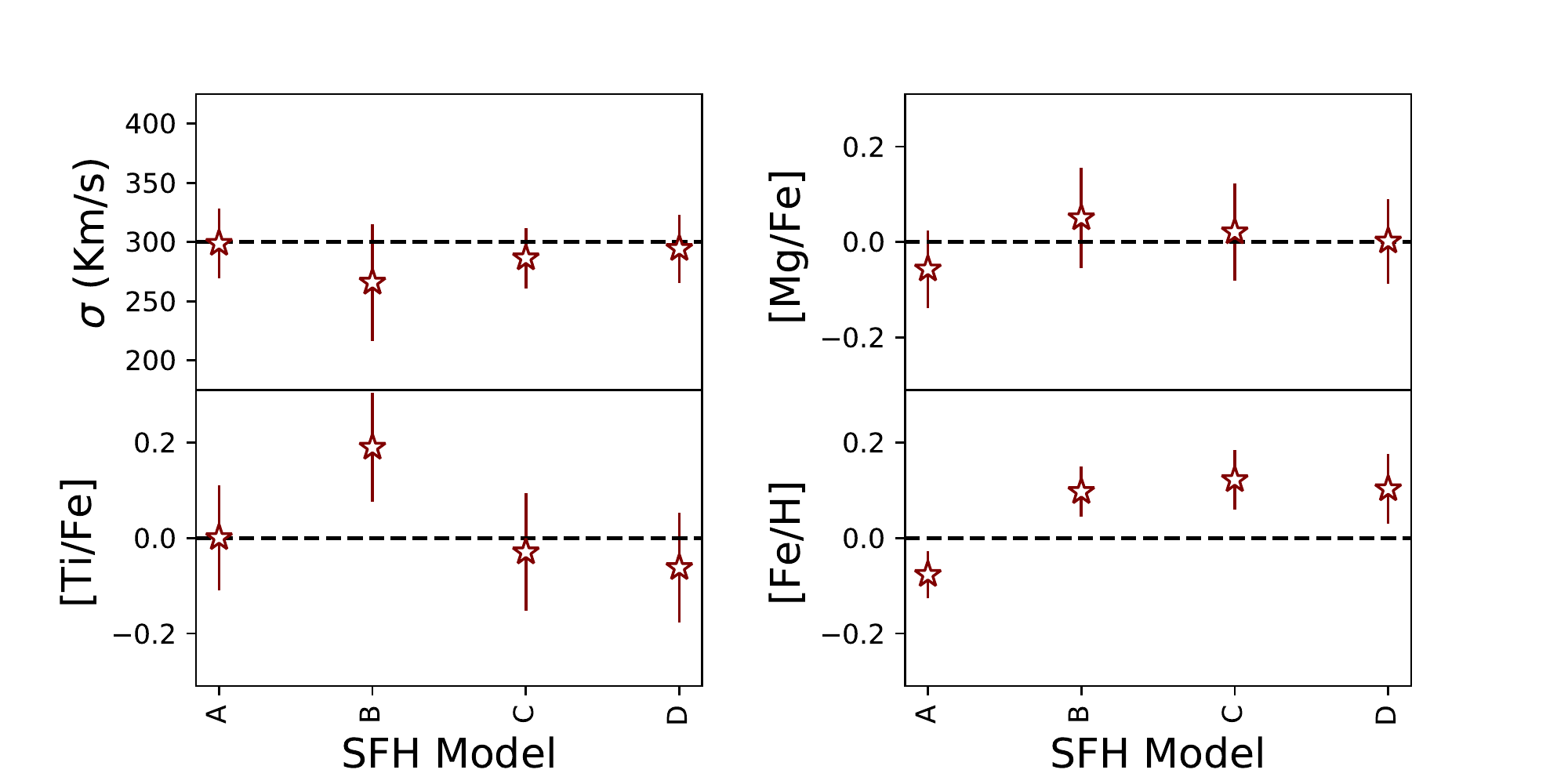}
\caption{The recovery of {\bf Top Left:} velocity dispersion and elemental abundances of {\bf Top Right:} [Mg/Fe], {\bf Lower Left:} [Ti/Fe], {\bf Lower Right:} [Fe/H]  from NIRSpec G235M/FL170LP grism/filter combination using {\tt alf}.
The models are generated using FSPS using different SFHs as detailed in Table \ref{tab:sfh_properties}. It is clear that with the exception of [Fe/H], other parameters are recovered accurately within the error limits for most SFHs.
\label{fig:alf_recoveries_FSPS}
}
\end{figure*}

In Figure \ref{fig:alf_recoveries_FSPS} we show the recovery of element abundances for the 4 FSPS mock observations. 
The [Mg/Fe] abundance of all 4 galaxies are recovered well by the {\tt alf} fits to our simulations. 
Similarly, except for model B, the [Ti/Fe] of the other SFHs are also well recovered. 
However, the [Fe/H] abundance cannot be recovered for any of the models. The [Fe/H] of the single burst (Model A) is underestimated by $\sim0.08$ dex, while the [Fe/H] of the other models are overestimated by $\sim0.1$ dex. 
Therefore, despite the fact that \citet{Villaume2017} SSPs showed good convergence of [Fe/H] for NIRSpec G235M/FL170LP S/N$\sim30$ observations, we find that FSPS models show a slight offset. 
We further run FSPS models up to S/N$\sim100$ and find that the discrepancy cannot be resolved by increasing the S/N of the observations and/or by a different grism/filter combination offered by NIRSpec. 
The underlying stellar libraries and isochrones used by {\tt alf} to fit the mock galaxy spectra are largely similar to the ones used to create the mock observables at $\lesssim 0.7\mu m$ ($\lesssim2.9\mu m$ in observed space at $z=3.2$). 
Thus, we speculate that the offsets we observe in Figure \ref{fig:alf_recoveries_FSPS} are due to effects such as dust and complex SFHs (except for Model A) in FSPS which are not accounted for in {\tt alf}. 
Further detailed analysis on the role of galaxy properties in the recoveries is warranted but is out of the scope of this paper.

\begin{figure*}
\includegraphics[ scale=0.36]{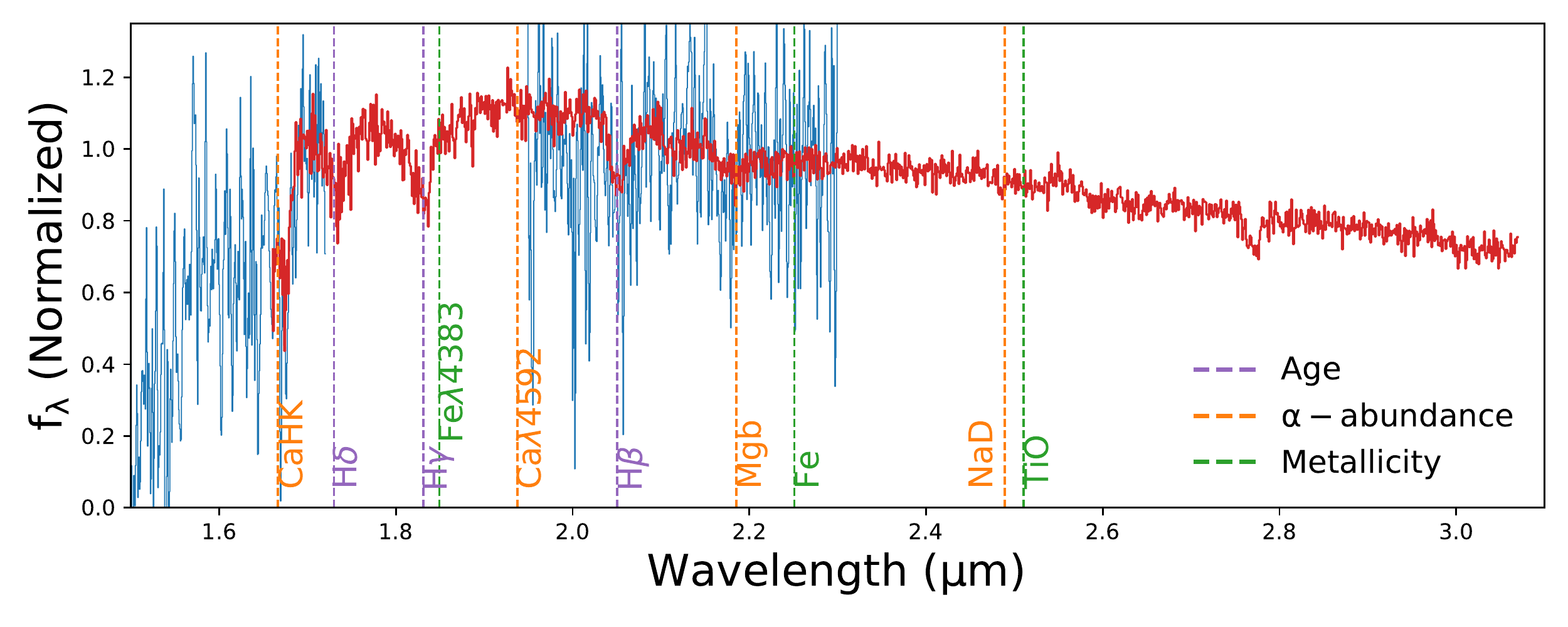}
\includegraphics[ scale=0.36]{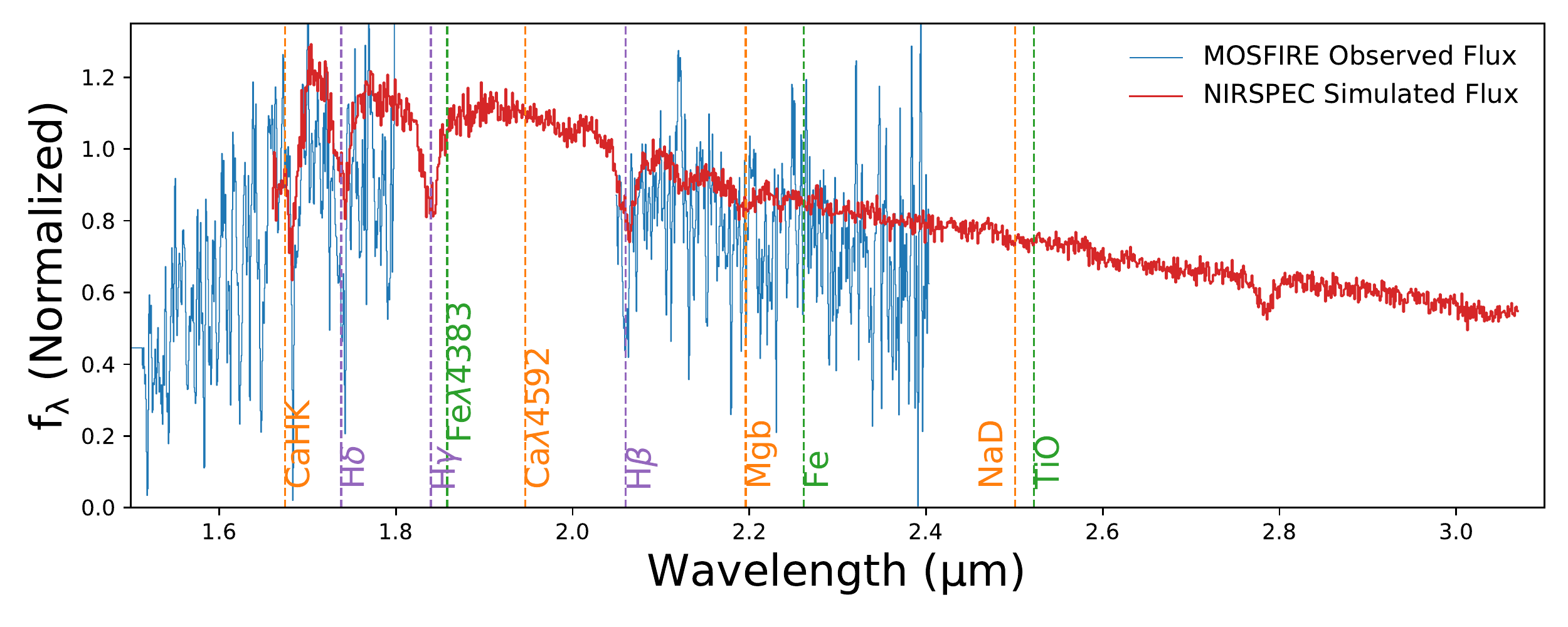}
\includegraphics[ scale=0.36]{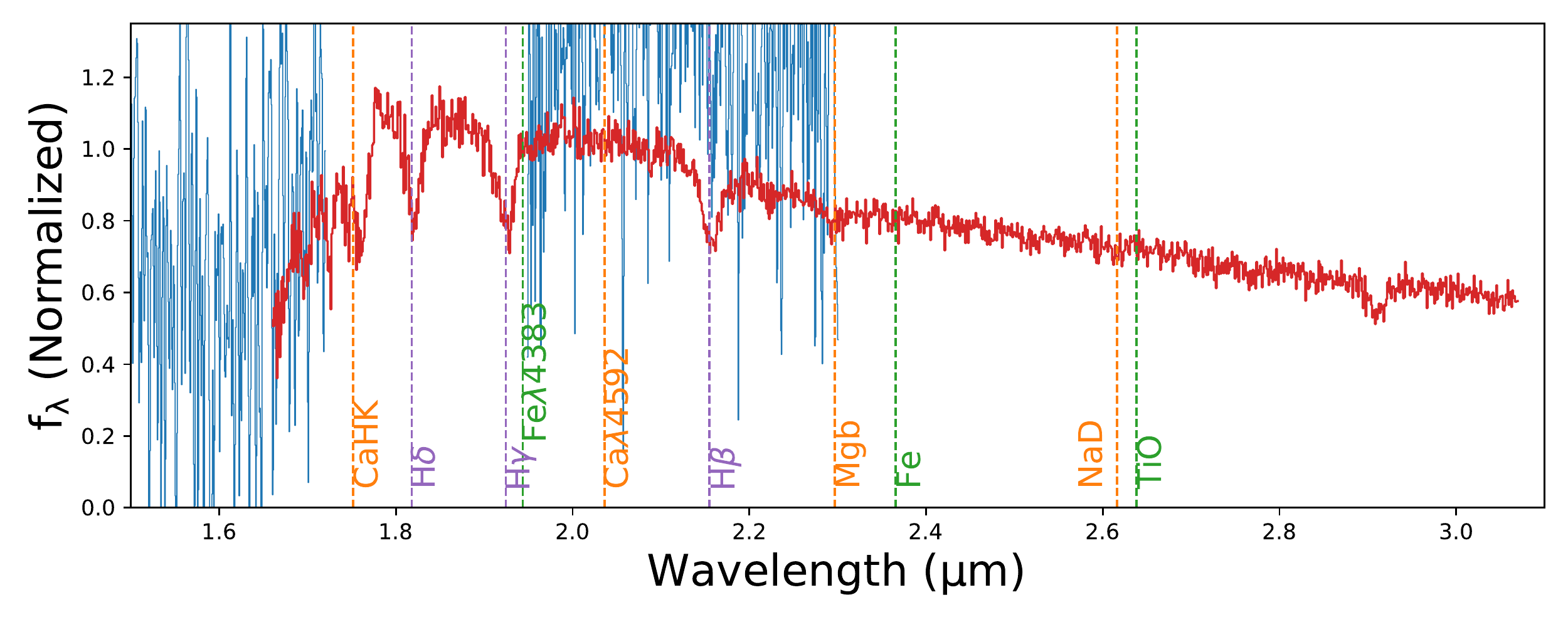}
\includegraphics[ scale=0.36]{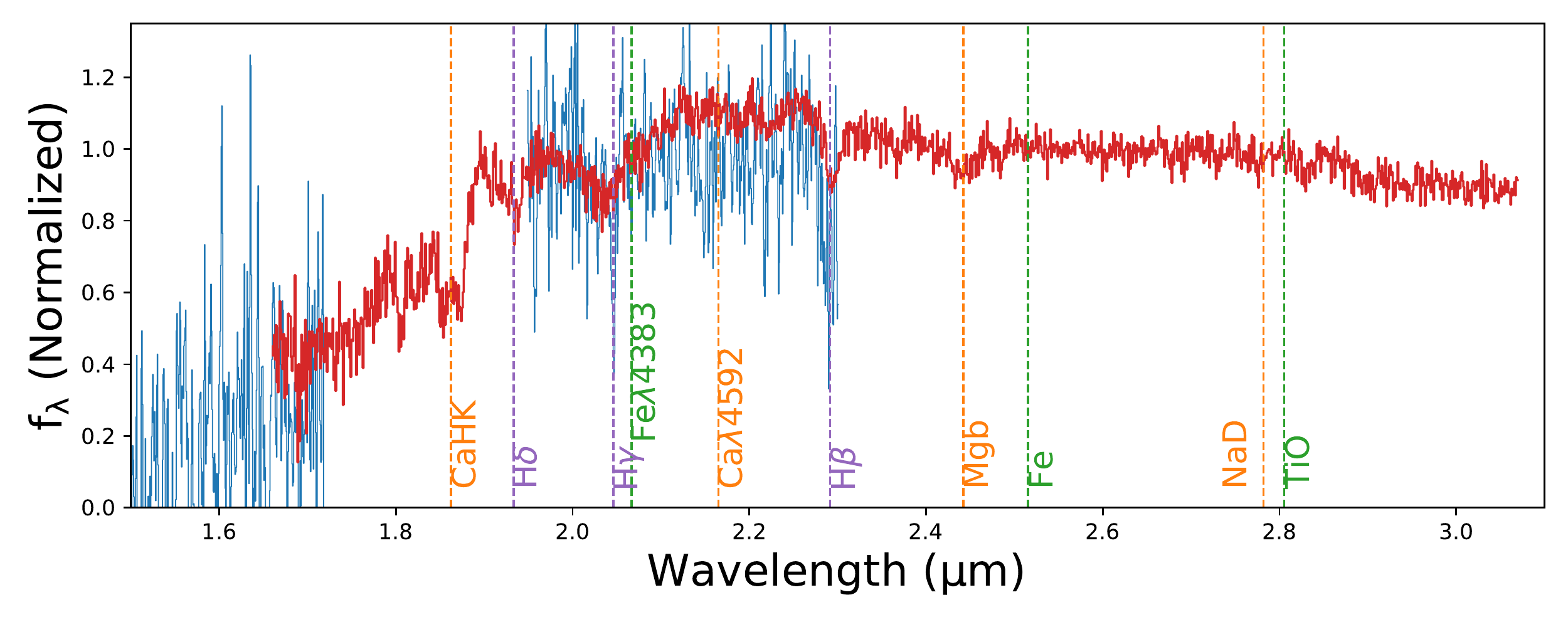}
\caption{Simulated JWST NIRSpec G235M/FL170LP observations of the four  $z\sim3-4$  quiescent galaxies presented by \citet{Esdaile2020}. The best fit \textsc{FAST++} templates to the galaxies from \citet{Schreiber2018} are used to obtain the JWST mock observations. We show {\bf Top Left:} 3D-EGS-40032 ($\sim8$h), {\bf Top Right:} 3D-EGS-18996 ($\sim4$h), {\bf Lower Left:} 3D-EGS-31322 ($\sim4$h), and {\bf Lower Right:} ZF-COSMOS-20115 ($\sim5$h) where the time stated inside the brackets refer to the typical NIRspec G235M/FL170LP observing times necessary to obtain a continuum S/N of $\sim30-40$. Ground based $H$ and $K$ band Keck/MOSFIRE spectra from \citet{Schreiber2018} are also shown for comparison. 
With $<10$ h of exposure time,  JWST/NIRspec can obtain $\gtrsim7$ times greater S/N quality at similar velocity resolutions compared to the current ground-based data for such targets and provides continuous coverage of features through atmospheric windows which are essential to analyze the stellar population properties of these galaxies. These features are colour-coded according to their primary sensitivity to age, $\alpha$-abundance, and metallicity.
\label{fig:simulated_spectra}
}
\end{figure*}

In Figure \ref{fig:simulated_spectra} we show the simulated JWST/NIRSpec G235M/F170LP observations of the four $z\sim3-4$ quiescent galaxies from the \citet{Esdaile2020} sample. These galaxies have per pixel S/N$\sim5-10$ in ground-based spectra with exposure times up to $\sim15$ h with Keck/MOSFIRE. 
Following morphological properties derived by \citet{Esdaile2020}, we reconstruct each source on the JWST ETC based on the HST WFC3/F160W imaging and use the best fit \textsc{ FAST++}\footnote{https://github.com/cschreib/fastpp} spectra from \citet{Schreiber2018} as the source spectra.

Following our aforementioned experiments with {\tt alf}, we alter the exposure times to obtain a continuum S/N$\sim30-40$, to maximize the accuracy of the recovered spectral parameters. 
With exposure times ranging between $4-8$ h, JWST/NIRSpec G235M/F170LP provide $\gtrsim7$ times better S/N at similar velocity resolutions compared to ground-based Keck/MOSFIRE spectra. 
It is also clear from the figure that, space-based observations from JWST allow a continuous coverage of most crucial rest-frame optical spectral signatures for the $z\sim3-4$ quiescent galaxy populations. These range from age, $\alpha$-element abundance, and overall metallicity sensitive indicators as shown by Figure \ref{fig:alf_features_abun}. 
Thus, rest-frame optical spectra of $z\sim3-4$ quiescent galaxies with similar properties to the \citet{Esdaile2020} sample can be obtained from JWST NIRSpec with modest exposure times. 
Such high-quality observations of quiescent galaxies are crucial to constrain their formation and evolution properties with high confidence.

\subsection{SFH}
\label{sec:sfh}

There is significant tension regarding the existence and abundance of massive quiescent galaxies in the $z>3$ universe between observations and cosmological simulation models \citep[e.g][]{Merlin2019a}. Thus, to understand the formation and subsequent quenching mechanisms of the early Universe, determining how massive quiescent galaxies built their mass efficiently is of great importance.

The star formation history of massive quiescent galaxies at $z\sim3-5$ could be addressed via two different methods.
Firstly, through analysis of individual line strengths such as Balmer and D4000 features, the approximate duration of the most recent SFH can be constrained \citep{Vazdekis2015a}. 
Additionally, if an enhancement of $\alpha$-elements such as Mg, Ca, and Na is observed, this would provide indications that they were formed over a short starburst event due to the delay time of Type Ia supernovae \citep[e.g][]{Kriek2016}.

Secondly, more stringent constraints to the SFHs can be made by combining full-spectrum fitting techniques with multi-wavelength photometric fitting techniques. 
However, even full spectrophotometric fitting techniques have degeneracies that result in a factor of $\sim2$ uncertainty in derived cosmological parameters, i.e. the cosmic star formation rate density \citep[e.g.][]{Madau2014,Yu2016b}. 
This is a result of multiple levels of complex degeneracies and uncertainties inherent to stellar populations models \citep{Conroy2013} and SED fitting techniques \citep{Walcher2011}. 
If only photometry is used in SED fitting, inferences are based purely on the SED shape. This can result in systematic offsets in derived galaxy properties due to effects such as the age-dust-metallicity degeneracy that affects the shape of the SED \citep{Bell2001}.

When spectra are combined with photometry, stronger inferences can be made on galaxy properties. 
This is due to extra information that can be obtained from the spectral features. 
However, uncertainties in stellar evolution, limitations in empirical stellar libraries, and limitations in theoretical stellar atmosphere models can introduce systematic biases to these inferences \citep[e.g.][]{Bruzual2007}. 
The SFH of a galaxy will have imprints in the overall shape of the SED and on the strength of different elements observed in the spectra. Additionally, the presence of specific types of stars leaves wider spectral imprints which can be used to provide further constraints \citep[e.g.][]{Brinchmann2008}. 
However, simultaneous modeling of a variety of spectral features that consider variations in stellar types, stellar/ISM abundances, and IMFs is complicated \citep{Gunawardhana2020}.  
When this is combined with non-parametric SFHs to  model the formation history, high-quality data and modular SED fitting codes become crucial \citep[e.g.][]{Leja2017,Leja2019a}.

Next, we perform mock observations to investigate whether spectrophotometric fitting of galaxies could recover the SFHs in sufficient accuracy to distinguish between different mass growth scenarios. 
Given the Universe is $\sim2$ billion years old at $z\sim3.2$, SFH variations should be evaluated by the SED fitting codes within this short time frame.  
We use the FSPS derived model galaxies with 4 different SFHs to generate the mock JWST NIRSpec spectra as detailed in  Section \ref{sec:stellar_pops}.
Observed photometry for the galaxies is computed using {\tt sedpy} python package mirroring the full wavelength coverage of the ZFOURGE COSMOS field \citep{Straatman2016}.
For simplicity, we assume that all photometric bands are detected at a S/N of 10.

Utilizing the mock photometry and deep NIRSpec rest-frame optical spectra with full-spectrum fitting techniques of \textsc{Prospector} \citep{Johnson2020} and its non-parametric approach to constraining the star formation histories \citep[e.g.][]{Leja2019a}, we investigate whether precise  constraints could be placed on the formation timescale of massive quiescent galaxies. 
In \textsc{Prospector} fitting, we allow stellar mass, stellar metallicity, and dust attenuation to vary as free parameters. 
Additionally, a {\tt continuity\_sfh} prior \citep[see Section 2.2.3 of][]{Leja2019a} is used for the SFH. 
In lookback time, the first two bins are fixed to be between 0-30 Myr and 30-100 Myr. The most distant bin is fixed to be between $0.85\times$age of the Universe and the age of the Universe, thus at $z=3.2$, the final bin is between 1700-2000 Myr in lookback time.
The remaining bins are split in logarithmic time evenly between 100 Myr and 1700 Myr.  
In total we use between 4-7 bins to investigate the SFH recovery for our 4 models.

\begin{figure*}
\includegraphics[ scale=0.42]{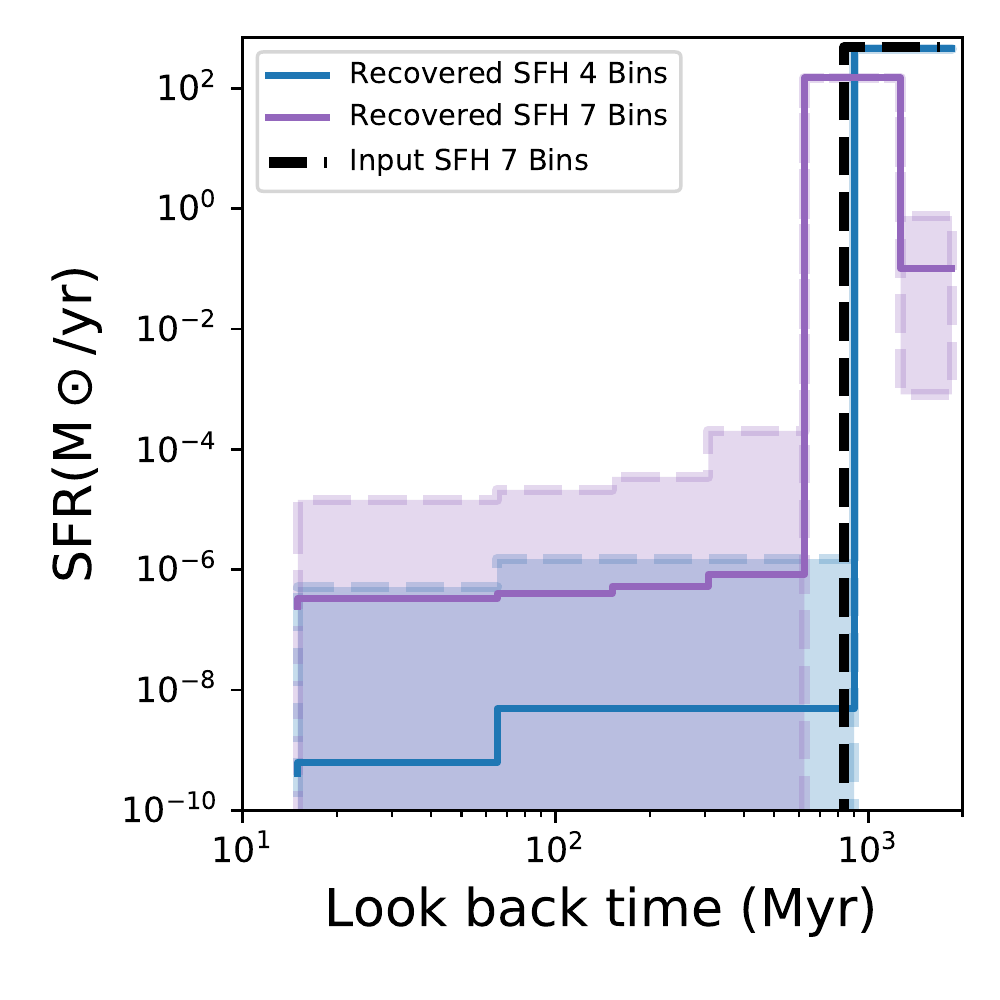}
\includegraphics[ scale=0.42]{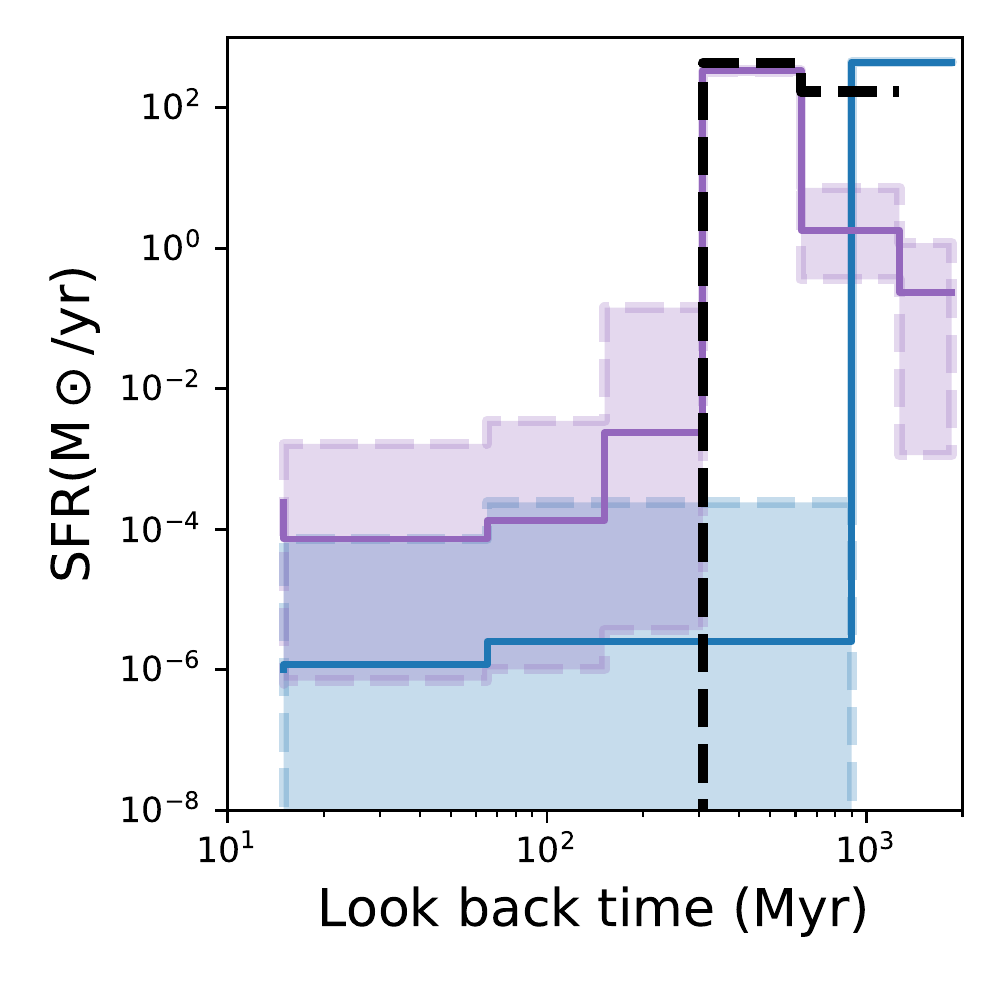}
\includegraphics[ scale=0.42]{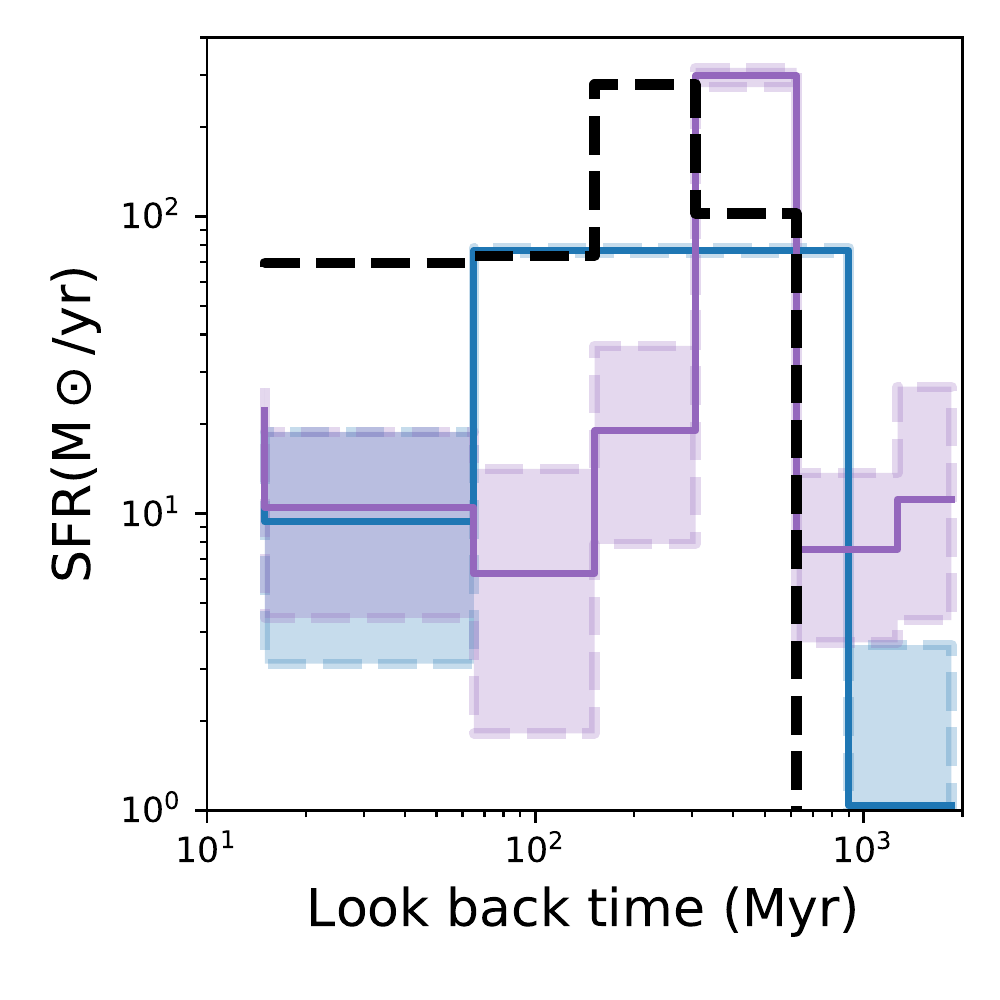}
\includegraphics[ scale=0.42]{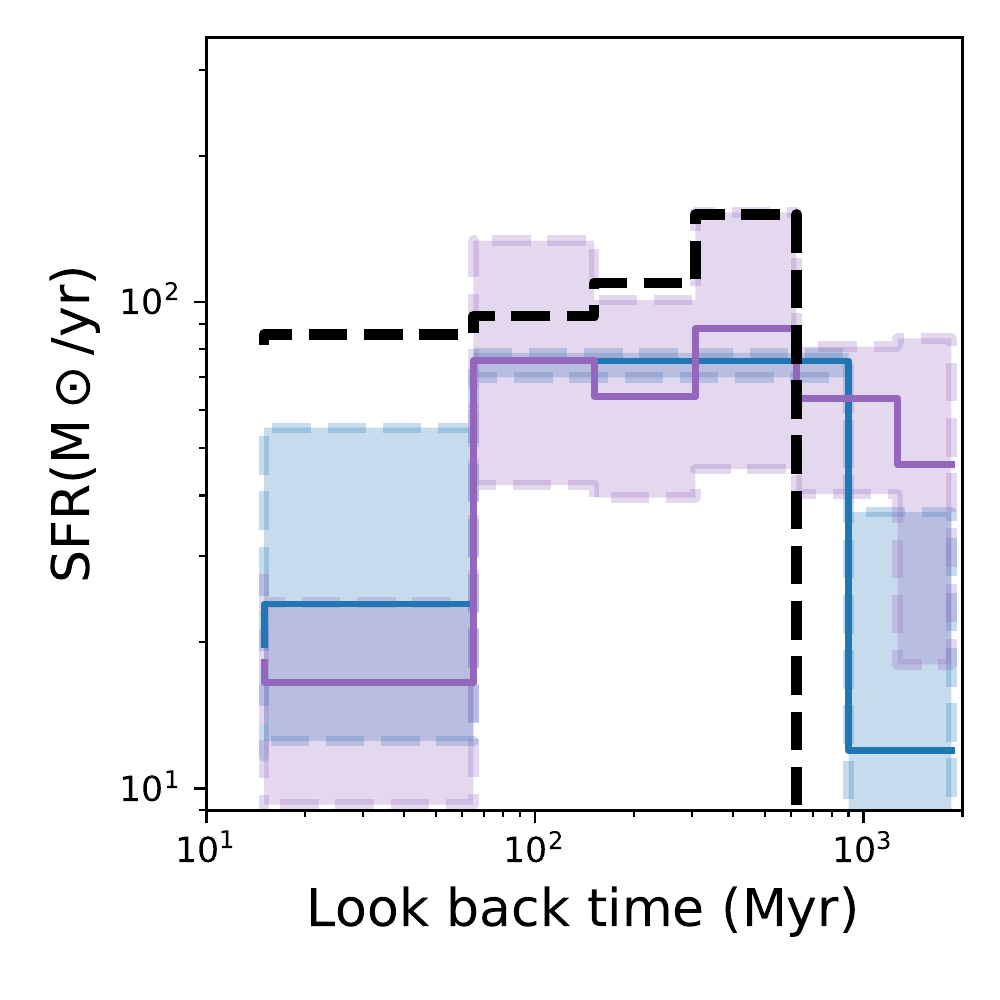}
\includegraphics[ scale=0.42]{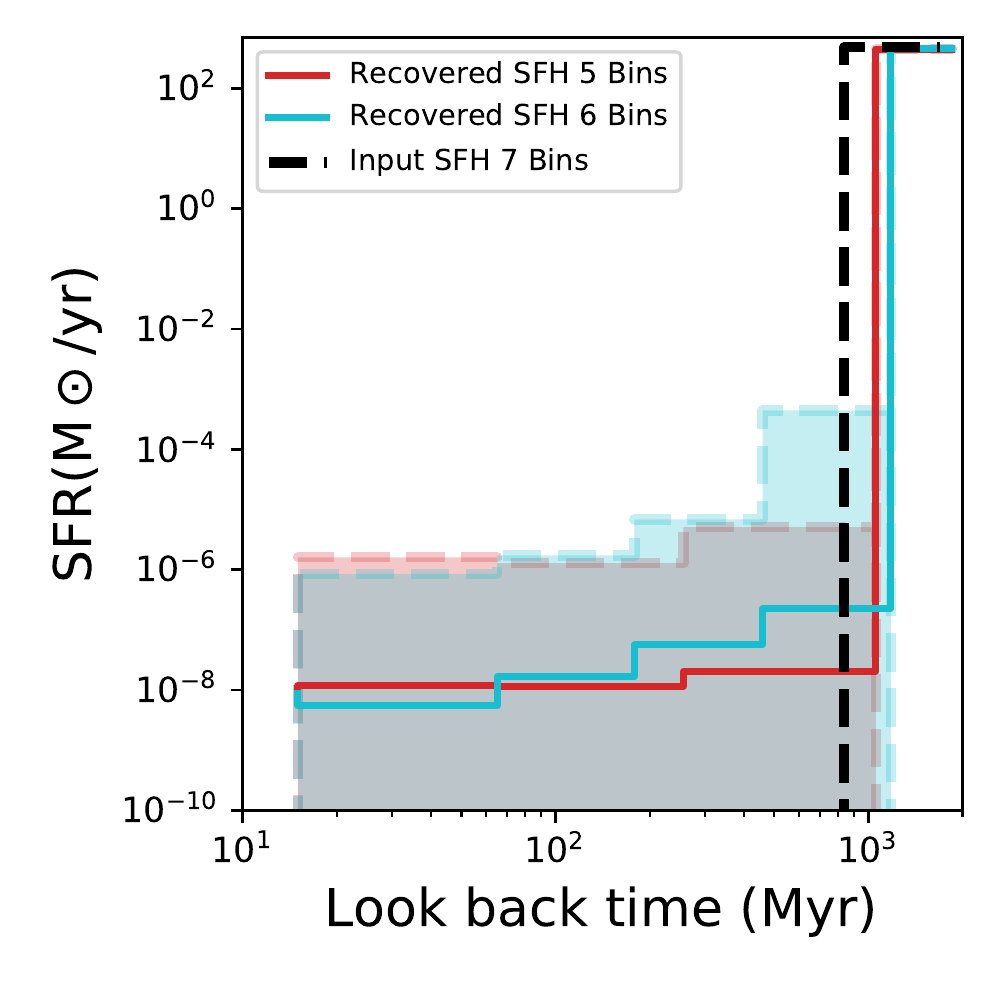}
\includegraphics[ scale=0.42]{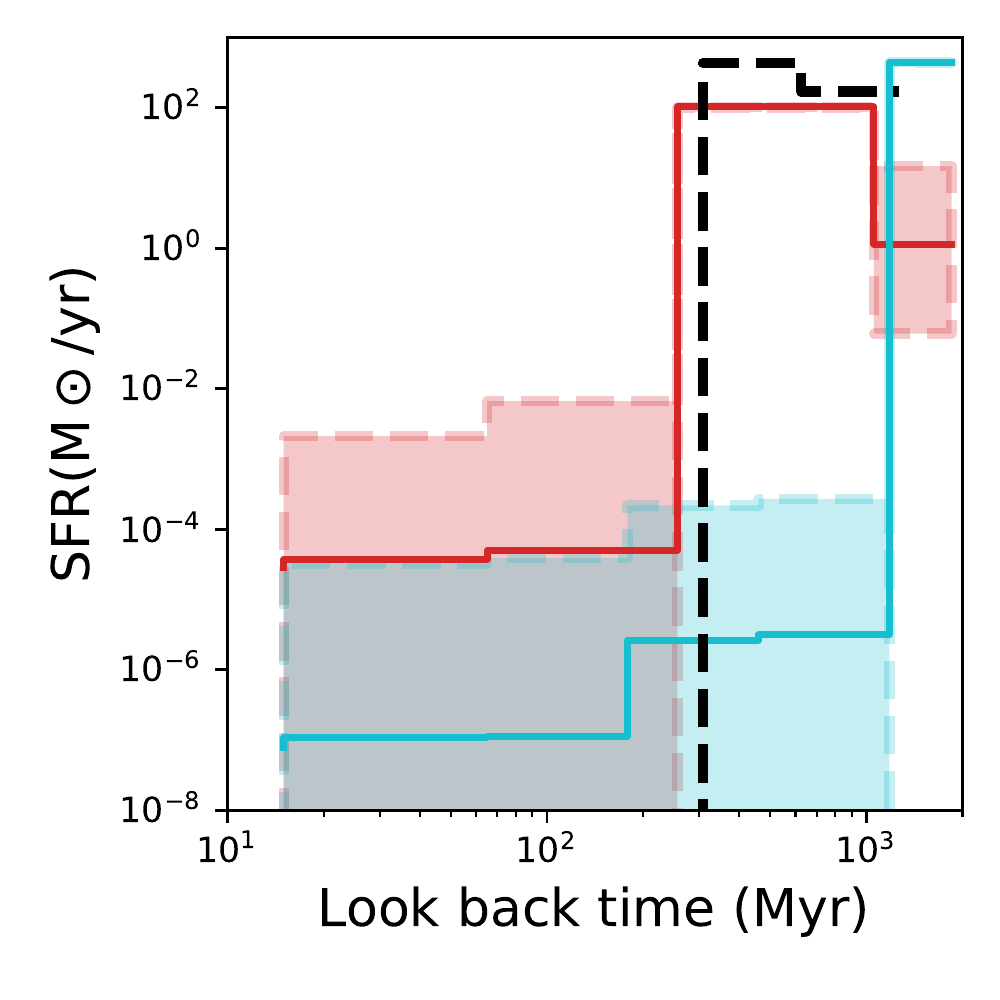}
\includegraphics[ scale=0.42]{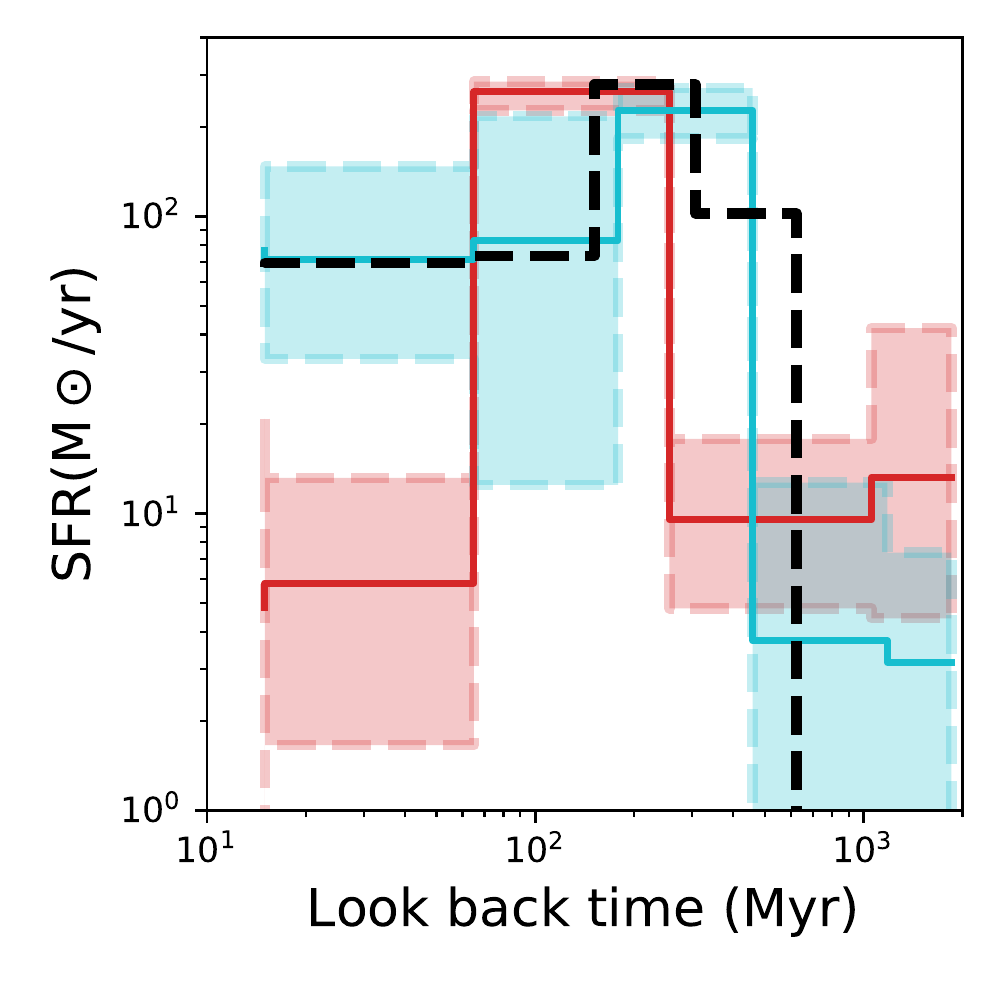}
\includegraphics[ scale=0.42]{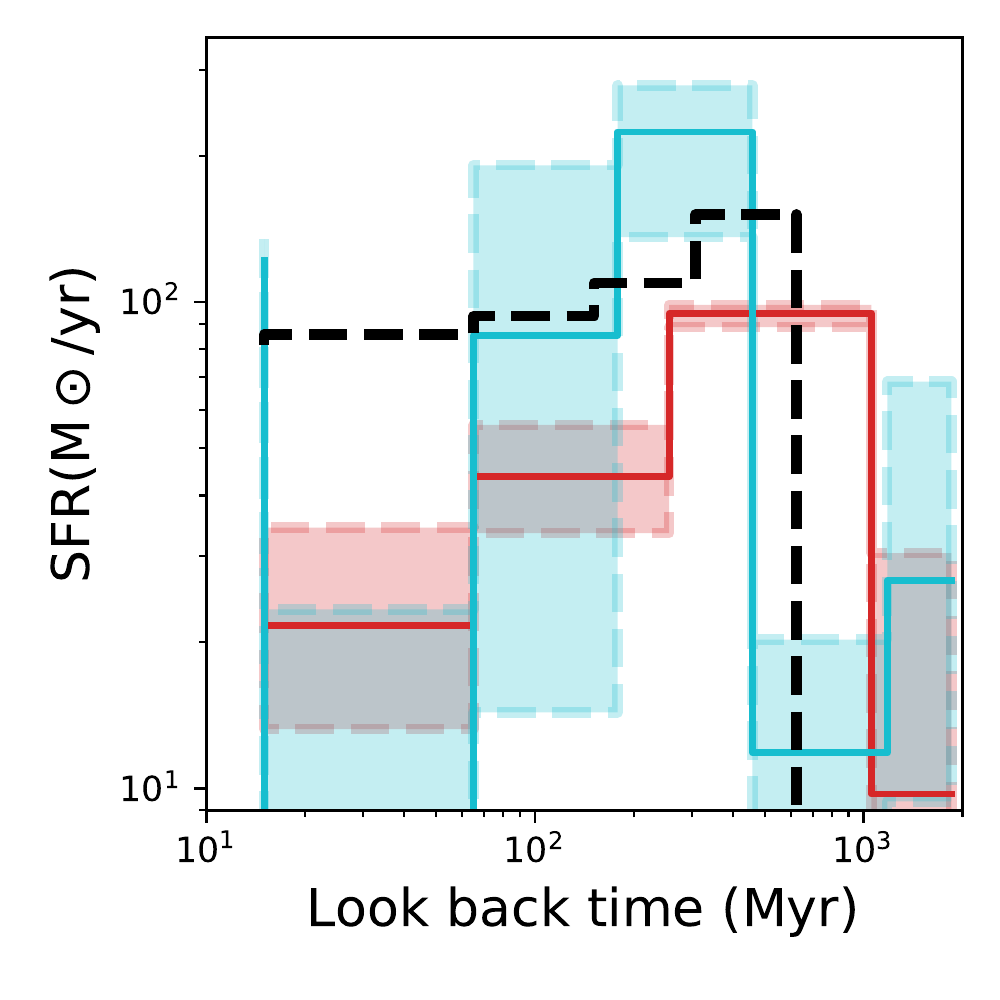}
\caption{The recovery of SFHs based on full spectral + photometry fitting by the \textsc{Prospector} SED fitting code. 
Panels show the recovery of Models {\bf First Column:} A {\bf Second Column:} B {\bf Third Column:} C {\bf Fourth Column:} D as described in Table \ref{tab:sfh_properties}. 
All models are fit with varying numbers of non-parametric bins ranging from 4-7 bins. The top row shows the recovery for 4 and 7 bins and the bottom row shows the recovery for 5 and 6 bins.
The input SFH is shown as a black dashed line using 7 time bins defined similarly to the 7 bin \textsc{ Prospector} fit. 
The 16th and 84th percentile are shown by the colour shading for each recovered SFH. 
For all models \textsc{Prospector} is able to recover the overall shape of the input SFH. However,  some fine tuning of the number of SFH bins is required to accurately recover the SFHs.  
\label{fig:sfh_recovery}
}
\end{figure*}

In Figure \ref{fig:sfh_recovery} we show the recovered SFH for our four models. 
For Model A, the form of the SFH is recovered well by \textsc{Prospector} independent of the number of SFH bins investigated here.
All recoveries suggest that the bulk of the star-formation happened around 1 Gyr and that the current SFR is very minimal. 
This is further confirmed by the highest time resolution 7-bin SFH showing a clear peak at $\sim1$ Gyr, with the SFR in the immediately preceding and following bins reducing by a $\gtrsim10^{-3}$ and $\gtrsim10^{-8}$ respectively.

For Model B, the increase in the SFR and the subsequent truncation of the SFR is well captured when the SFH is modeled using either 5 or 7 bins. The 4 and 6 bin SFHs fail to recognize the exponential increase in the SFR at earlier times. 
This is expected for the 4-bin SFH, because the 3rd time bin by construct covers a large 100-1700 Myr time window. The galaxy only forms stars for 500 Myr in this window, which may result in the SFH sensitivity of this bin to reduce. However, the 6-bin SFH should have had sufficient sensitivity to recover the increase in SFR.  
The lack of star-formation in the later stages of the galaxy is well recovered irrespective of the number of bins.

For Model C, the input SFH shown by the black dashed line is recovered well by all the fits. 6 and 7 bin SFHs provide the most accurate recovery of the burst, however, even with the limited time resolution of 4 and 5 SFH bin fits, the burst is still captured by the \textsc{Prospector} fits. 
The 7 and 5 bin SFHs bins, also show that the SFR was declining before the galaxy underwent a burst. This, however, is not captured by the SFH with 6 bins even though it has finer time sampling compared to the SFH with 5 bins.  
It is interesting to note that at $z\sim3.2$, even with 7 time-bins, the input SFH does not show the exponentially declining nature of the SFH in the post burst phase (compared to Figure \ref{fig:input_sfhs} model C). Thus, constraining the full form of the SFH would require a much larger number of time-bins. 
Additionally, only the 6 bin SFH is able to detect the ongoing star-formation of this model at present time.

For Model D, the overall shape of the SFH is captured well by the fits. 
The 6 bin SFH suggests the galaxy to have a younger $\sim500$ Myr component and an older $\sim2$ Gyr component. 
The 7-bin SFH also suggests that a significant fraction of the galaxy is been formed at an older age compared to the input model. 
None of the recoveries are able to capture the residual star-formation of this model at present time.

When comparing the recoveries between the models, it is evident that \textsc{Prospector} struggles to recover the ongoing star-formation for Models C and D (with the exception of the 6 bin SFH recovery for Model C). 
Therefore, even with state-of-the-art SED fitting codes like \textsc{Prospector} , it is plausible that galaxies with ongoing residual star-formation could  be identified as quiescent when a full spectrophotometric fitting is performed. 
Analysis of optical rest-frame nebular optical emission lines from JWST spectroscopy could be used to provide constraints on the star-formation limits of such galaxies. 
However, low emission lines such as \OII\ have been observed in lower redshift quiescent galaxies \citep[e.g.][]{Maseda2021}, which adds extra complexity. 
Dust continuum detections in the sub-mm would provide an independent constraint to the SFRs in such galaxies \cite[e.g.][]{Simpson2017,Schreiber2018b}.

Albeit limitations in recovering residual star-formation at later times, in general, our simulations with different SFHs and the different number of SFH bins demonstrate that the SED fitting techniques (investigated here in the context of \textsc{Prospector}) can recover the past form of the SFH to distinguish between different formation scenarios (extended formation vs short sharp formation). 
We note that our simulations here are tied closely to investigate the recovery of parameters for galaxies that are of similar properties to the \citet{Esdaile2020} sample. 
In this context, in order to recover the input SFHs accurately, some tuning of the number of SFH bins is required. 
The differences between bins are likely due to the sensitivity of certain photometric/spectral features to the SFH which are enhanced with finer/coarse time sampling. 
However, these should be well tested with simulated models for galaxies with short evolutionary times (age $\rm \lesssim 3$ Gyr) and compared with other recovered galaxy properties such as stellar mass and metallicity.  
We defer this to a future analysis. 
We also note that the rest-frame optical includes both age and abundance-sensitive lines, thus both the SFH and metallicity can be constrained to high accuracy using \textsc{Prospector} with fine-tuned binning. Additionally, when direct $\alpha$-element constraints are lacking, metallicity along with strong constraints on the SFH can be used to infer the $\alpha$-abundances.

\subsection{The IMF}

The stellar initial mass function is a fundamental parameter in galaxy evolution that plays a vital role in regulating galaxy formation and chemical evolution of the Universe. Even though traditionally the IMF has been considered to be universal \citep[e.g.][]{Salpeter1955,Kroupa2001,Baldry2003,Chabrier2003}, recent results have started to show evidence for systemic variations \citep[e.g.][]{Bastian2010,Hopkins2018}. 
At $z\sim0$ studies have shown that the lower mass of the IMF slope could  vary as a function of velocity dispersion \citep[e.g.][]{Cappellari2012,Conroy2013b}, metallicity \citep[e.g][]{Navarro2015c}, and radial distance from the galactic center \citep[e.g.][]{Navarro2015a} in local early type galaxies (ETGs). 
However, mechanisms for such variations are still not well understood and overall agreement between different IMF indicators are still not clear \citep[e.g][]{Smith2014,Smith2020}. 
Constraining the IMF in the early Universe is vital to determine how such variations may have been influenced by conditions prevalent in the Universe when the local ETGs were building up their stellar masses.  
Constraints on the early Universe IMF is also necessary to provide constraints to the cosmic SFH and to understand how the Universe transformed from hydrogen and helium to the current complex Universe.

The IMF of the $z\sim3-5$ massive quiescent galaxies can be constrained using high-precision dynamical and stellar mass measurements.  NIRSpec slit spectroscopy or  NIRSpec IFU observations (see Section \ref{sec:dynamics}) can be used to constrain the dynamical masses of the galaxies. Given these early systems are compact and dense (M$_*\sim10^{10}-10^{11}$, $\rm r_e\sim0.5-2 kpc$ \citep{Straatman2014,Esdaile2020}), the baryonic matter is expected to dominate the kinematics, thus dynamical mass measurements are largely independent of the assumed dark matter halo profiles. 
This means that the dynamical mass is only contributed to by the stellar mass and thus the ratio of the dynamical mass to stellar mass (also known as the IMF mismatch parameter \citep[e.g.][]{Davis2017}) can be used to understand the stellar populations via the IMF.
JWST NIRSpec spectroscopy will also provide constraints to the rest-frame optical spectral shape of the $z\sim3-5$ massive quiescent galaxies, which will provide strong constraints to stellar masses \citep{Conroy2013}. 
Therefore, with JWST spectroscopy, stronger constraints on the IMF mismatch parameter can be obtained.

The IMF mismatch parameter can be compared with the distribution of the $z\sim3-5$ galaxies in the stellar mass plane to see if there are differences in the inherent stellar populations of these galaxies, perhaps indicating different formation scenarios. 
Given quiescent galaxies by definition do not have active star-formation, the `observed' dynamical mass will only constrain the lower mass slope of the IMF (but is degenerate with high mass stellar remnants). This is due to the lack of massive O, B type stars, which have relatively short lifetimes of $10-100$ Myr compared to the long-lived lower mass stars.

The IMF mismatch parameter is the only practical way to measure the IMF of unlensed individual massive quiescent galaxies at $z\sim3-5$. 
As we showed in Figure \ref{fig:alf_features_abun}, IMF sensitive spectral features such as the Na I doublet and the Wing–Ford band resulting from K and M dwarf stars require S/N $>300$ to distinguish between different IMF slopes \citep{vanDokkum2010}. 
Our NIRSpec simulations with {\tt alf} show that a S/N$\sim350$ is required in the G395H/FL290LP grism/filter combination to recover the IMF slopes and reaching such levels even for the brightest $z\sim3-5$ massive quiescent galaxies is not feasible due to the very high exposure times required.

Apart from the `observed' IMF, the \emph{relic} IMF of $z\sim3-5$ massive quiescent galaxies can be reconstructed through spectrophotometric fitting of advanced SED fitting codes such as \textsc{Prospector} (private communication) and variable IMF chemical evolution codes such as {\tt galIMF} \citep{Yan2019}. 
By providing IMF as an extra free parameter in SED fitting, the evolution of the IMF in the $z>5$ Universe can be constrained. 
However, the time variation of the IMF, SFH, and $\alpha-$element abundances are intricately connected \citep[e.g][]{Navarro2016}, therefore detailed stellar population modeling is required to understand degeneracies between them. 
$z>5$ Universe is the time window when the $z\sim3-5$ massive quiescent galaxies were forming the bulk of their stars \citep[e.g][]{Valentino2020}. 
If these galaxies did indeed have top-heavy IMFs in their star-formation phase (e.g. as formulated by \citet{Lacey2016}, also see \citet{Gunawardhana2011,Nanayakkara2017,Nanayakkara2020}), this would lead to differences in the number of ionizing photons produced, the chemical enrichment. and stellar wind and supernovae feedback processes. 
Such changes would be crucial to the re-ionization timescales of the Universe, thus, the reconstruction of the \emph{relic} IMF in the $z>6$ Universe is important to constrain the cosmological evolutionary models of the Universe.


\section{Dynamical Properties of galaxies with JWST}
\label{sec:dynamics}

The dynamical properties of massive $z\sim3-5$ quiescent galaxies are an important quantity to constrain the evolutionary properties of massive galaxies in the early Universe.  
Through JWST slit or IFU spectroscopy, velocity dispersions, dynamical masses, stellar masses, sizes, and the evolution of the mass-size-velocity dispersion plane of quiescent galaxies at $z\sim3-5$ can be constrained. 
Analyzing the $z\sim3-5$ galaxies in this plane and comparing them with galaxies at lower redshifts is necessary to build up the cosmic evolutionary picture of quiescent galaxies and to determine how they evolve in the mass-size plane with cosmic time \citep[e.g.][]{Belli2017a}. 
Constraining kinematic properties is also important to determine whether these massive quiescent galaxies have undergone mergers in their evolutionary history \citep[e.g.][]{Belli2017b}, which will shed further light into mass assembly and quenching processes in the early Universe. 


\subsection{Velocity dispersions through slit spectroscopy}
\label{sec:dynamics_slit}

JWST NIRSpec slit spectroscopy can be used to measure the velocity dispersions of $z\sim3-5$ massive quiescent galaxies. In Figures \ref{fig:alf_recoveries_GXXXH} and \ref{fig:alf_recoveries_G235} we show the recovery of the velocity dispersions of our mock NIRspec S200A1 simulations. It is clear that with both the grisms and filters, the velocity dispersion is converged to the input value of the mock simulations. In Section \ref{sec:abundances} we established that G235M/FL170LP grism/filter combination provides the best wavelength coverage and the necessary resolution to obtain the element abundances of $z\sim3-5$ galaxies.

In terms of velocity dispersion, the G235M medium-resolution disperser of NIRSpec can obtain dispersions of $\sim130$ km/s for a 2-pixel resolution element while  the high-resolution G235H disperser achieves a dispersion resolution of $\sim50$ km/s. 
Given the lowest velocity dispersion observed for  $z>3$ quiescent galaxies is $\sim150$ km/s \citep{Esdaile2020}, the medium resolution grating is sufficient to obtain velocity dispersions of quiescent galaxies. 
We also perform mock JWST observations of a galaxy with $\sim150$ km/s velocity dispersion and find that at a S/N of 30, the velocity dispersion can be recovered by the G235M/FL170LP grism/filter combination. 
Additionally, we find that the delivered S/N from G235M is higher than the S/N of G235H when binned to the same spectral resolution.

By obtaining a S/N of $\sim30$ per pixel with JWST NIRSpec G235M/FL170LP observations, robust constraints can be obtained to the dynamical masses. For example, \citet{Esdaile2020} measured dynamical masses for four $z\sim3-5$ massive quiescent galaxies to an accuracy of $\sim20\%$. This was based on measurements of Balmer and Ca H \& K absorption lines covering a total of $\sim600$ \AA\ in rest-frame with a continuum S/N of $\sim5-7$ achieved through spectral binning of 6 \AA\ in the rest-frame.
With NIRSpec, our simulations show that high-quality spectra of such galaxies can be obtained with typical S/N of $\sim30-40$ per pixel at similar resolution with continuous coverage of $\sim3000$ \AA\ in the rest-frame. 
They have no skyline contamination and reach an accuracy of $\sim2-3\%$ for dynamical mass measurements. 
This higher level of accuracy will enable robust comparisons with mass-matched control samples of galaxies at $z\sim2$, which can also be obtained as filler targets using the JWST NIRSpec multi-object spectroscopy (MOS) which utilizes the NIRSpec micro-shutter assembly (MSA).


\subsection{Galaxy kinematics through IFU spectroscopy}
\label{sec:kinematics_IFU}

\begin{figure*}
\includegraphics[ scale=1.0]{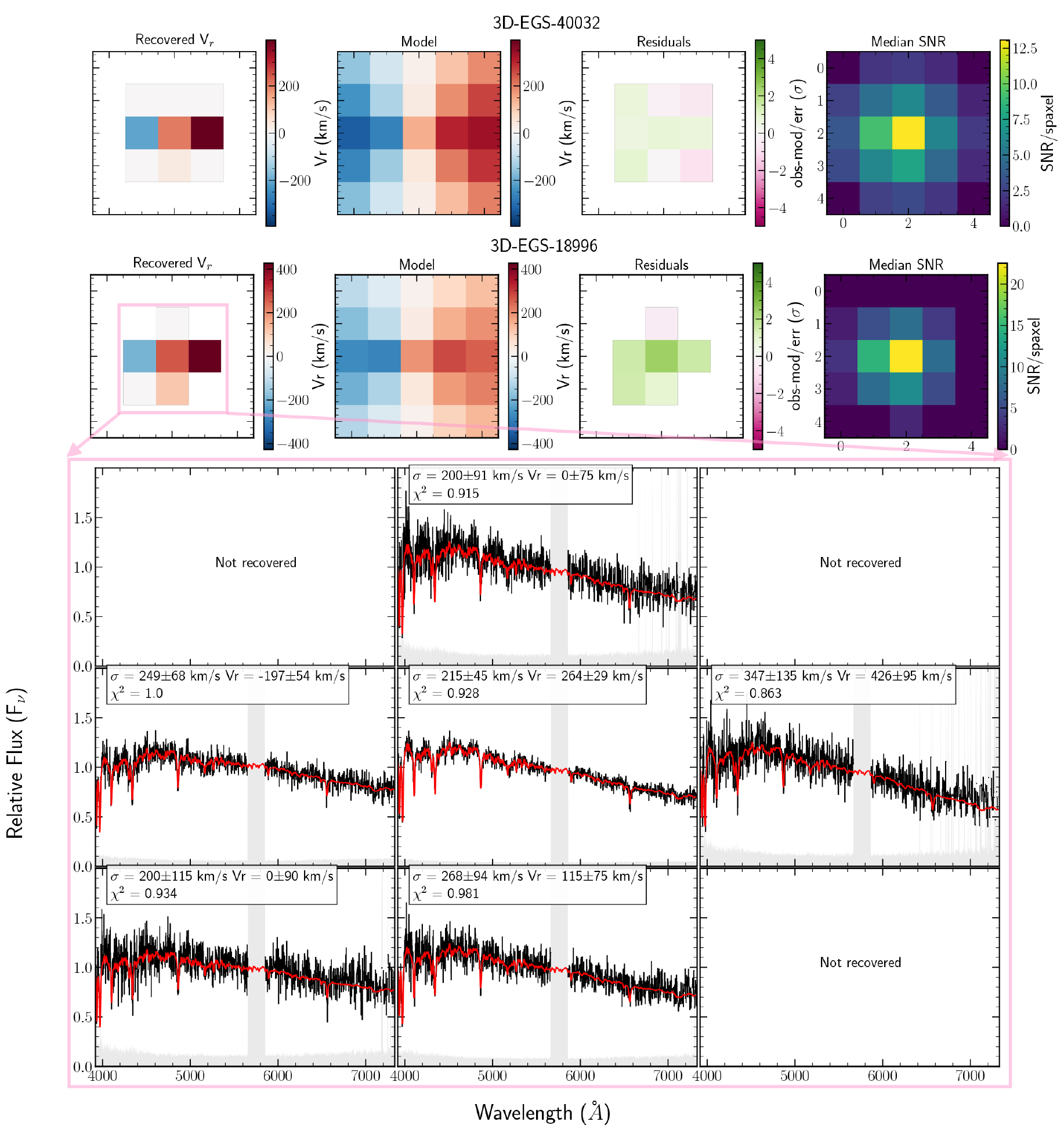}
\caption{Simulated IFU observations for 2D kinematics recovery of two galaxies in the \citet{Esdaile2020} sample; 3D-EGS-40032 (G140M/F100LP grism/filter, 12h exposure) and 3D-EGS-18996 (G235H/F170LP grism/filter, 8h exposure). The chosen grism/filter combination optimizes the spectral ranges (which includes [OII] emission from 3D-EGS-40032), sensitivity, and resolving power to measure existing velocity dispersions for each galaxy.
{\bf Top: from left to right:} recovered 2D kinematics from simulated IFU cube{ \bf (left)} for a Hernquist 2D model of a rotationally supported galaxy and with an inclination of 45\degree {\bf (middle-left}), the normalized residuals {\bf (middle-right)} and median S/N/spaxel from the IFU cube. The simulation for 3D-EGS-40032 is in the top row with a $V_r/\sigma=1.5$ and 3D-EGS-18996 is below with a $V_r/\sigma=1.8$. {\bf Bottom:} Recovered radial velocities and velocity dispersions for 3D-EGS-18996 per inner set of spaxels in the simulated IFU observations cube. Black lines are the simulated spectra, red lines are the best-fit and gray shaded area is the corresponding noise spectra. The spectral fit is only included for $ 0.85 < \chi^2 < 1.5$
\label{fig:ifu_sim}
}
\end{figure*}

JWST NIRSpec IFU spectroscopy can be used to obtain resolved kinematics of $z\sim3-5$ massive quiescent galaxies.  
Compared to ground-based IFU spectroscopy, the JWST NIRSpec IFU provides many advantages for kinematics analysis. 
Higher spatial resolution can be obtained due to diffraction-limited space-based observing at $2\ \mu m$ and is not limited by the lack of bright stars necessary for adaptive optics corrections to increase the seeing. 
Generally, non-AO observations are preferred for the analysis of continuum sources with absorption-line kinematics.
This is because, in kinematics, the AO PSF imposes an interdependency between luminosity, velocity, and dispersion, thus more advanced deconvolution is necessary for each spectral plane of the original data cube \citep{Davies2012}. 
This can be corrected by robustly using emission line features (after accounting for beam smearing), however, for continuum features a PSF convolved step-wise model for every pixel in the continuum and equivalent width would be required. 
This is hard to achieve even for bright sources \citep{Thater2019}.    
Additionally, the desired S/N is practically impossible to achieve even with the best ground-based IFUs such as Keck/OSIRIS, where the  majority of the absorption features for $z\sim3$ galaxies exist in the $H-$band where the Strehl ratio is low ($\sim10\%$). 
For example, for the two mock NIRSpec IFU simulations of two of the \citet{Esdaile2020} galaxies which we describe below, the total exposure time necessary to obtain resolved kinematics increases from $\sim8-12$ hours with JWST/NIRSpec time to $>40-100$ hours (based on the surface brightness of the galaxy) in ground-based Keck/OSIRIS spectroscopy.

In order to investigate whether JWST NIRSpec IFU observations could recover the kinematics of $z\sim3-5$ massive quiescent galaxies accurately, we perform full 2D mock IFU simulations to investigate the recovery of spatially resolved velocities with pPXF \citep{Cappellari2017}. 
We use empirical SSPs from \citet{Villaume2017} to generate mock 1.5 Gyr old galaxies with velocity dispersions consistent with the measurements in \citet{Esdaile2020}. 
These model spectra are input into the JWST ETC, including their physical size and properties determined from modeling using $HST$ $F160W$ band images. Several mock IFU cubes were then generated to obtain a suite of S/N values. Hernquist rotational models \citep{Hernquist1990} were then applied to each simulated IFU observation, assuming a 45\degree inclination, with various rotational-to-velocity dispersion ratios ($V_r/\sigma$) to simulate galaxies with rotational support. 
A Hernquist model is chosen as suitable for modeling a rotating spheroidal star system; the estimated mass and the scale length (taken to be the effective radius of the galaxy), produce the velocity profile. The Plummer model \citep{Plummer1911} produces similar rotational curves. 
Finally, we used pPXF on the instrument calibrated data to fit for radial velocity and velocity dispersion.

In Figure \ref{fig:ifu_sim} we show the recovery of both 3D-EGS-18996 and 3D-EGS-40032 2D kinematics. We note that the JWST ETC outputs 0.1'' spaxels which do not account for drizzling, however, we expect in practice that drizzling will improve the sampling compared to what is shown here. Given the compact sizes and sharp decrease in S/N per spaxel, binning to this IFU spaxel scale is likely optimal for signal, while still allowing sufficient spatial scale to infer rotation.
To assess the S/N requirements for the mock observations, this analysis was repeated for different exposure times. We found that a median S/N $\sim7$ per spaxel is required in the adjacent to central pixels to recover a radial velocity and infer rotation with $>2\sigma$ confidence. At this S/N $V_r/\sigma$ can be constrained to $V_r/\sigma \sim 1.5$ in 3D-EGS-18996 and $V_r/\sigma \sim1.3$ in 3D-EGS-40032 and confirm rotational support in these galaxies. Additionally, this would limit the underestimation in dynamical mass from 0.4 dex (for a $V_r/\sigma \sim3$) to $\sim0.1-0.15$ dex \citep[from equation 5][]{Belli2017a} in both galaxies and allow for high precision IMF constraints without potential large systematic errors.


\section{Discussion and Conclusions}
\label{sec:discussion}

Through a case study of the \citet{Esdaile2020} sample, we have shown here that the launch of JWST will address crucial observational challenges that currently limit our understanding of the abundance, formation, and evolutionary mechanisms of $z\sim3-5$ quiescent galaxies. Given these galaxies provide  insights into galaxy evolution in the $z>6$ Universe, the detailed analysis would determine how the first generation of galaxies in the Universe formed and evolved. 
Current cosmological models find it challenging to match the observed abundance of massive quiescent galaxies in the $z\sim3-5$ Universe \citep[e.g.][]{Merlin2019a,Valentino2020}, thus the analysis of the stellar populations and galaxy kinematics is vital to reconstruct the star formation and merger history of these galaxies.

The development of medium-band imaging instruments has allowed surveys like ZFOURGE \citep{Straatman2016} and FENIKS \citep{Esdaile2021} to be carried out where the photometric redshifts can be obtained with high accuracy. This is driven by the  increased spectral sampling of the split $H$, $J$, and $K$ bands \citep[e.g.][]{Nanayakkara2016}. 
JWST NIRSpec CLEAR spectroscopy provides the most prominent avenue to probe the completeness of these photometrically selected massive quiescent galaxies in the early Universe. This is driven by the multiplexing capability of NIRSpec along with the simultaneous continuous wavelength coverage between $0.6-5.3\ \mu m$. 
The large wavelength coverage is well suited to identify star-forming contaminants in $z\sim3-5$ quiescent candidates. 
In Figure \ref{fig:prism_spectra} we showed that the resolution and the sensitivity of NIRSpec PRISM/CLEAR spectroscopy was sufficient to accurately distinguish between quiescent, post star-burst, and dusty star-forming galaxies. 
However, the spectral resolution of PRISM/CLEAR spectroscopy is not sufficient to perform a detailed analysis of element abundances to constrain the stellar population properties of $z\sim3-5$ massive quiescent galaxies.

The optimal strategy to obtain spectral features to perform a detailed analysis of stellar populations is to obtain deeper high-resolution JWST NIRSpec spectroscopy for samples that are already confirmed to be quiescent and are at a redshift range of interest. 
Using simulated spectra from \citet{Villaume2017}, we generated mock NIRSpec observations for G235M/FL170LP, G235H/FL170LP, and G395H/FL290LP grism/filter combinations to identify the best grism/filter combinations and the S/N levels required to recover accurate element abundances and velocity dispersions. 
We found that a S/N of $\sim30$ is required to reach the necessary accuracy levels for element recoveries and that the G235M/FL170LP grism/filter combination is optimally suited to carry out the necessary observations. We found that the lower resolution of the G235M enables more efficient observations by achieving the necessary S/N level without compromising the accuracy, even when the higher resolution G235H grism is binned to the same velocity. 
Further analysis with parametric SFHs from FSPS \citep{Conroy2010b} showed that the G235M/FL170LP grism/filter combination could still largely recover element abundances and velocity dispersions. 
However, we found small offsets between input and recovered [Fe/H] abundances for these parametric SFH models.

The G235M/FL170LP grism/filter combination is ideal to obtain coverage of rest-frame optical features of the $z\sim3-5$ massive quiescent candidates. As an example we showed that \citet{Esdaile2020} sample will cover the crucial age, $\alpha$-abundance, and metallicity sensitive indicators, all within $<8$ h of exposure time. Compared to current best ground-based spectroscopy, the improvement on S/N from JWST NIRSpec was found to be $7\times$ at similar velocity resolutions. 
As we showed in Figure \ref{fig:alf_features_abun}, the Balmer absorption features are most sensitive to the ages, Mgb is strongly sensitive to the $\alpha$ abundance, and the overall metallicity is well constrained from the various Fe features that fall within the G235M/FL170LP wavelength coverage. With combined full-spectrum fitting with software like {\tt alf}, the stellar population properties can be constrained to high accuracy.

In addition to the most recent star-formation episode that can be recovered from the Balmer absorption features, novel advanced SED fitting codes such as \textsc{Prospector} \citep{Johnson2020} allow the full star-formation history of galaxies to be recovered. 
Using mock JWST G235M/FL170LP observations and photometry mapping the ZFOURGE COSMOS field coverage, we showed that spectrophotometric fitting  can accurately recover the SFHs of galaxies to distinguish between different formation mechanisms. 
However, we observed the accuracy of the outcome to be dependent on the number of SFH bins and that \textsc{Prospector} struggled to recover the ongoing star-formation for some of the models.
More work on understanding the role of the number and duration of SFH bins in accurately recovering the SFH of galaxies with short formation timescales is warranted.

Future advancements of SED fitting codes will allow IMF to be varied as an extra free parameter (i.e. \textsc{Prospector}, private communication). This will enable astronomers to perform robust statistical modeling of SFHs with variable IMFs in the early Universe. 
Recent studies have suggested that star-forming galaxies are likely to have a higher fraction of high mass stars compared to the traditional \citet{Salpeter1955} IMF \citep[e.g.][]{Gunawardhana2011,Nanayakkara2017} and recent semi-analytical models have started to implement such changes to their evolutionary models \citep{Lacey2016}. 
However, adding extra layers of free parameters could lead to larger degeneracies between the parameters. 
Additionally, with the superior quality NIRSpec data and advanced spectrophotometric fitting codes these extra degeneracies will be further constrained by the confident detections of multiple $\alpha$-elements, Fe, and the suite of Balmer absorption lines. They  will provide stringent constraints to metallicity, element abundances, and SFH timescales.

Apart from the reconstruction of the SFHs, the integral field spectrographic capabilities of JWST/NIRSpec allow the kinematics of the $z\sim3-5$ massive quiescent galaxies to be explored. 
In Figure \ref{fig:ifu_sim} we showed through mock observations that the kinematics of galaxies similar to \citet{Esdaile2020} sample can be recovered. 
In addition to constraints to dynamical masses (and inferences about stellar population properties such as IMF), the kinematics of such galaxies provide vital clues to whether these massive galaxies already show ordered rotation or if they show signatures of mergers.
This can be used as an independent method to establish the formation pathways to early massive quiescent galaxies. 
Thus, future surveys with JWST will be able to determine whether the early massive quiescent galaxies could be formed within the first $\sim1-2$ Gyr following $\Lambda$CDM hierarchical merger cosmology.

With the advancements made in understanding the abundances, properties of the stellar populations, and formation scenarios of $z\sim3-5$ massive quiescent galaxies through JWST, the next generation of cosmological simulations will be able to address the cosmic puzzle of how such galaxies in the early Universe were so efficient in building up stellar mass within short time-frames. 
This will allow additional constraints on galaxy evolution in the $z>6$ Universe to be made, thus will provide more stringent constraints on reionization pathways and timescales of the Universe and the relative contribution of massive galaxies to the epoch of reionization.


\section*{Acknowledgments}
We thank Michael Maseda, Marijn Franx, and Joel Leja for helpful discussions. 
We also thank the staff at the James Webb Space Telescope Help Desk. 
T.N., K. G., M.D., and C.J. acknowledge support from Australian Research Council Laureate Fellowship FL180100060. 
This project made use of {\tt astropy} \citep{Astropy2018}, {\tt matplotlib} \citep{Hunter2007}, and {\tt pandas}.




\bibliographystyle{mnras}

\bibliography{bibliography.bib} 







\end{document}